\documentclass[12pt]{article}    

\input epsf.tex

\newcommand{\be}{\begin{equation}}
\newcommand{\ee}{\end{equation}}
\newcommand{\bea}{\begin{eqnarray}}
\newcommand{\eea}{\end{eqnarray}}

\newcommand{\gs}{\ensuremath{g_s}}      
\newcommand{\ls}{\ensuremath{l_s}}      
\newcommand{\ms}{\ensuremath{m_s}}      

\newcommand{\npb}[3]{Nucl.~Phys. {\bf B#1} (#2) #3}

\def\p{\partial}
\newcommand{\tr}{\mathop{\rm Tr}}

\def\brack#1{\langle #1 \rangle}
\newcommand{\cO}{{\mathcal{O}}}
\newcommand{\cN}{{\mathcal{N}}}

\newcommand{\cL}{{\mathcal{L}}}
\newcommand{\bS}{{\mathbf{S}}}

\newcommand{\cT}{{\mathcal{T}}}

\newcommand{\Ophi}{{\mathcal O}_{\phi}}

\newcommand{\Ima}{{\mathrm{Im}}}
\newcommand{\rl}{r_{\Lambda}}
\newcommand{\zl}{z_{\Lambda}}
\newcommand{\Z}{\zeta}
\newcommand{\Zl}{\zeta_{\Lambda}}
\newcommand{\X}{\xi}
\newcommand{\vm}{{\mathbf{m}}} 
\newcommand{\rv}{\rho}  
\newcommand{\cNl}{{\mathcal{N}_l}}  
\newcommand{\CC}{\sum_{\vm}C^{9\ldots 9}_{l\vm}C^{9\ldots 9}_{l\vm}} 
\newcommand{\vy}{\vec{y}}
\newcommand{\hy}{\hat{y}}
\newcommand{\vq}{\vec{q}}
\newcommand{\vk}{\vec{k}}

\begin{document}
    
\begin{titlepage}

\begin{flushright}
UUITP-02/00\\
USITP-00-05\\
hep-th/0004187
\end{flushright}

\vspace{1cm}

\begin{center}
{\huge\bf D3-brane Holography}
\end{center}
\vspace{5mm}

\begin{center}
    
{\large Ulf H.\ Danielsson,$^{\scriptstyle 1}$
Alberto G\"uijosa,$^{\scriptstyle 2}$ \\
\vspace{3mm}
Mart\'\i n Kruczenski,$^{\scriptstyle 1}$ 
and Bo Sundborg$^{\scriptstyle 2}$}\\

\vspace{5mm}

$^{1}$ Institutionen f\"or Teoretisk Fysik, Box 803, SE-751 08 
Uppsala, Sweden

\vspace{3mm}

$^{2}$ Institute of Theoretical Physics, Box 6730, SE-113 85 
Stockholm, Sweden

\vspace{5mm}

{\tt
ulf@teorfys.uu.se, alberto@physto.se, \\
martink@teorfys.uu.se, bo@physto.se
}

\end{center}

\vspace{5mm}

\begin{center}
{\large \bf Abstract}
\end{center}
\noindent

It has recently been conjectured that the AdS$_{5}$/SYM$_{4}$ correspondence 
can be generalized 
away from the conformal limit, to a duality between supergravity on the full
asymptotically flat 
three-brane background and a theory characterized as $\cN=4$ SYM deformed in 
the
IR by a specific 
dimension-eight operator. Assuming that this relation is valid, we derive a
prescription for computing 
$n$-point correlation functions in the holographic theory, which 
reduces to the
standard AdS/CFT recipe 
at low energies.  
One- and two-point functions are discussed in detail.
The prescription follows from very simple considerations and
appears to be applicable to any asymptotically flat background.
We also compute the quark-antiquark 
potential and comment on the description of the baryon in the 
supergravity picture.
We conclude with some comments on the possible relation between our 
work and recent results in non-commutative field theories.

\vfill
\begin{flushleft}
April 2000
\end{flushleft}
\end{titlepage}
\newpage
    
\section{Introduction}

The path that led to the discovery of the AdS/CFT 
correspondence\footnote{See
\cite{magoo} for 
a comprehensive review and an extensive list of references.} 
\cite{malda, gkp,w} began with the 
realization that the physics of D-branes can be captured from two quite distinct
perspectives: as extended objects in
supergravity, or as localized objects with intrinsic worldvolume dynamics
\cite{polchrr}. Over the years, an enormous body of evidence has
accumulated in support of this insight. In the particular case of
D3-branes, the works 
\cite{3/4,klebabs,gkt,gk}
performed a variety of comparisons between quantities in the two alternative
descriptions, and constituted 
very important steps on our way to unraveling the precise relation between the
two approaches.
It was from these works that Maldacena \cite{malda} was able to distill his
statement of equivalence between the extreme
low-energy limit of the worldvolume dynamics (governed by $\cN=4$ SYM in four
dimensions) and type IIB string 
theory in the near-horizon region of the three-brane supergravity solution 
(i.e., $\mbox{AdS}_{5}\times\bS^{5}$). 

More recently, the authors of \cite{ghkk,gh,intri} have attempted to take
another step along this path, by
exploring the possibility of elevating the AdS/CFT correspondence to a duality
between type IIB string theory
defined on the full asymptotically flat three-brane background, and the
effective theory describing the 
low-energy worldvolume dynamics of D3-branes at strong 't~Hooft coupling. The
latter theory has been characterized 
in \cite{gh,intri} as an IR deformation of the $\cN=4$ fixed point by a specific 
dimension-eight operator.   To directly examine this non-renormalizable gauge
theory would constitute
an enormously difficult challenge. 
In this paper, we pursue a different line of
attack. We assume that there exists a holographic image of physics
on the D3 background, even if its precise form may be unknown to us.
By definition,
all information about this holographic dual is encoded in the bulk 
supergravity theory, since the duality mapping must work 
in both directions. 
Our approach will thus be to study the duality
by developing a calculational
procedure that extracts this information from
the bulk theory. 

The paper is structured as follows. 
In the next section we give a more detailed
account of the
proposal of \cite{ghkk,gh,intri}, and delineate our approach.
In Section \ref{2ptsec} we derive a
prescription for computing two-point
correlators, using as a concrete example
the operators that couple to dilaton partial waves.
Section \ref{abssec} contains a 
discussion of
the two-point function we obtain, demonstrating that it 
reduces to the known $\cN=4$ SYM result at low energies, and 
establishing its
relation with the supergravity
absorption probability. The result differs in
some aspects from the
one discussed by previous authors \cite{ghkk,gh}. 
In Section \ref{1ptsec} we calculate a one-point function
in the presence of an external source,
and use it to discuss the
UV/IR relation, encountering some interesting features.
The prescription is then
generalized to arbitrary $n$-point functions
in Section \ref{nptsec}. 
Sections \ref{wilsonsec} and \ref{baryonsec}
are respectively devoted to a calculation of 
the quark-antiquark potential and the
description of baryons
as D5-branes.
A final section presents our conclusions, and includes some comments 
on the close parallel between certain aspects of our results and recent 
work in non-commutative field theories.
 
\section{The Physics of D3-branes} \label{backgroundsec}

In this section we will carefully examine the regimes of applicability 
of the two alternative descriptions of a system of D3-branes, 
and review the proposal of
\cite{ghkk,gh,intri} for generalizing the AdS/CFT correspondence 
\cite{malda} to a duality involving the full D3-brane background.
We should emphasize from the outset that,
although our work was motivated by the existence of this relatively 
concrete duality conjecture, the route we choose to follow does not 
explicitly rely on its specific form. We will 
simply assume there exists some brane
theory dual to supergravity on the 
three-brane background, and proceed in Sections 3 and onward to 
derive rules for computing quantities in the dual theory in terms of 
physics on the curved background. 

\subsection{Dual descriptions of D3-branes}

Consider a collection of a large number, $N$, of D3-branes
in Type IIB string theory. 
On the one hand, the physics of this system can be
described in terms of a worldvolume theory coupled to
string theory in the bulk of flat ten-dimensional spacetime.
For processes with substringy energies, 
$\omega\ll \ms$, only the lowest 
modes of the D3-branes and the closed strings
can be excited, and the description is in terms 
of a non-Abelian gauge theory coupled to supergravity. 
The action for the combined system is then
\be 
S=S_{\mathrm{S}}\left[\phi,h,\ldots\right]+
  S_{D3}\left[A,\Phi,\ldots;\phi,h\ldots\right]~,
\ee
where the two terms
denote the Type IIB supergravity action in a flat background and 
the D3-brane action, respectively. The degrees 
of freedom on the branes
include a gauge field $A^{\mu}_{ab}$, six scalars 
$\Phi^{i}_{ab}$ ($a,b=1,\ldots,N$; $\mu=0,\ldots,3$; $i=4,\ldots,9$),
and their fermionic superpartners.
For discussion purposes
it is convenient to split the 
D3-brane 
action\footnote{See \cite{taylorvr2,myers} and
references therein for an account of what 
is known about the form of D$p$-brane 
low-energy effective actions.}
into two terms,
$S_{D3}=S_{\mathrm{b}}[A,\Phi,\ldots]+
S_{\mathrm{int}}[A,\Phi,\ldots;\phi,h,\ldots]$,
describing the dynamics on the brane and the couplings to 
the supergravity fields, respectively.
The brane action
$S_{b}$ includes in particular
contributions from the Born-Infeld term,
which are schematically of the form
\be \label{bi}
S_{\mathrm{b}}=
    -N\int d^{4}x\,\tr \left\{ F^{2}+R^{4}F^{4}+\ldots\right\}~.
\ee
Note that the gauge field $A$,
as well as the scalars $\Phi$ (not shown 
explicitly), have been
't~Hooft-normalized. This is the 
normalization which is most convenient in studying the large $N$ limit.

On the other hand, the system under consideration can be studied as a 
black brane solution of supergravity, with metric
\bea \label{d3metric}
ds^2 &=& H^{-1/2} (-dt^2 +dx_1^{2} + dx_2^{2} +dx_3^2 ) 
+   H^{1/2} (dr^2 +r^2 d\Omega_{5}^2),  \\
H(r)&=& 1+{R^{4}\over r^{4}}, \qquad 
R^{4}=4\pi N\gs\ls^{4}~, \nonumber
\eea
a constant dilaton $e^{\bar{\phi}}=\gs~$, 
and $N$ units of Ramond-Ramond flux
through the five-sphere. 
The above metric describes a geometry with
an asymptotically flat region $r\ge R$, 
and a throat extending from $r= R$
down to a horizon at $r=0$. 
One can trust this supergravity 
solution as long as $R\gg \ls$,
or in other words, $N\gs \gg 1$.

Henceforth we will refer to the above two descriptions
as the flat and the curved pictures, respectively.
It should be emphasized
that simultaneous validity of the two descriptions
requires
\be \label{restriction}
\omega\ls\ll 1, \qquad R/\ls\gg 1,  
\ee
but
the combination $\omega R$ is arbitrary,
as pointed out already in \cite{klebabs}.
When this combination is small it represents a convenient
expansion parameter.

That
the curved and flat space pictures should be in some sense equivalent
has been clear
since Polchinski's identification of D-branes as 
RR-charged black branes \cite{polchrr}.
At the string level
one has the option of \emph{either} formulating the worldsheet theory 
as a non-linear $\sigma$-model for closed strings
in the non-trivial three-brane background, \emph{or} introducing 
explicit D3-branes by considering holes on the worldsheet
with appropriate boundary conditions. We emphasize that in the latter 
description the spacetime metric one perturbs about is 
flat\footnote{Notice that
there is no inconsistency here--- even if the number of
D-branes is large, the worldsheet theory with Dirichlet/Neumann boundary 
conditions and
a flat target-space metric is certainly conformal.
To introduce D-branes as boundary conditions \emph{and at the same
time} consider the associated
non-trivial background would be double-counting.}, the 
non-trivial 
geometry having being traded for the cumulative effect of
open string loops. 

Initial evidence for the equivalence of the two pictures
came from 
the early D-brane scattering 
calculations \cite{dscatt}. 
A more 
systematic exploration of the precise connection between
the two descriptions
began with the comparison of thermodynamic quantities 
\cite{3/4} and
absorption cross-sections
\cite{klebabs,gkt,gk} 
computed in the two approaches. 

Based partly on these works, Maldacena motivated his duality 
conjecture \cite{malda} by noting that
in the low-energy 
limit,
$\omega R\to 0$, the 
branes in the flat picture
decouple from the bulk (i.e., $S_{\mathrm{int}}\to 0$). 
In addition, it 
can be seen from (\ref{bi}) that in 
this limit only the leading term survives, and the worldvolume theory 
reduces to $\cN=4$ SYM.  In taking the limit, the dimensionless
ratio $\Phi/\omega$ 
should be held fixed. It is convenient and
customary to regard $\omega$ as fixed and finite, in which case the 
low-energy decoupling is achieved by 
taking\footnote{In \cite{malda} this limit was written  
as $\ls\to 0$, which is equivalent to $R\to 0$, since
$R/\ls\propto (g_{s}N)^{1/4}$ is held fixed.} $R\to 0$.

The expectation value of the Higgs field is related to the radial 
coordinate in the curved 
picture through $r=R^{2}\Phi$. So if $R\to 0$ with $\Phi$ fixed,
we have $r\propto R^{2}\to 0$,
and as a result $H(r)\to (R/r)^{4}$, which means that we zoom in on  
the near-horizon geometry, AdS$_{5}\times\bS^{5}$.
The gauge theory energy $\omega$ corresponds in
the curved picture
to the energy measured at infinity, and is related to
the locally measured energy $\omega _{r}$ through 
$\omega=[H(r)]^{-1/4}\omega_{r}$.
So even though $\omega R\to 0$,
the fact that $r/R^{2}$ is held fixed
implies that $\omega _{r}R$ stays fixed and finite, and consequently
stringy excitations remain in the spectrum.
Maldacena's remarkable conclusion is that
the full string theory in AdS$_{5}\times\bS^{5}$ can be equated
with $\cN=4$ SYM \cite{malda}. 
The string and SYM couplings are related
through $g_{YM}^{2}=2\pi\gs$. 
For $\gs\ll 1$ and $\gs N \gg 1$ the bulk description is
in terms of classical supergravity. 
Quantum $\gs$ and stringy $\ls/R$  corrections about this limit
are mapped onto  $1/N$ and
$1/\lambda^{1/4}$ corrections in the gauge theory,
respectively, where $\lambda=g_{YM}^{2}N$ is the 
't Hooft coupling.

Let us now try to obtain a holographic dual that describes 
more than just
the near-horizon geometry. 
Such information would be included if we keep $\omega R$ 
finite
\cite{klebabs,ghkk,dealwis}.
It is clear from (\ref{bi})
that  $S_{b}$ will then include terms of dimension higher than four.
The worldvolume theory is thus no longer conformally 
invariant, its behaviour
at different scales being described by a particular
renormalization group trajectory. 
Just like in the AdS case, we expect a 
UV/IR correspondence
\cite{susswi,pp} to operate, relating
the bulk radial coordinate $r$ to an energy
scale in the field theory. 
The
non-conformal nature
of the worldvolume theory
is thus simply a reflection of
the fact that the supergravity background 
is no longer $SO(4,2)$-invariant, and as a result it has 
different properties 
at different values of $r$. 

In accord with (\ref{restriction}), we of course still restrict
attention to substringy energies. But
because we now
work away from the extreme low-energy limit, 
the branes and the bulk do not decouple. As pointed out
in \cite{gh}, we can still achieve a decoupling of 
sorts in
the weak coupling limit\footnote{In Section \ref{2ptsec}
we will be more precise about the sense in which the
branes `decouple' from the bulk in this limit.}
$\gs\to 0$. The flat space supergravity 
theory becomes free, and
interactions can only take place on the 
worldvolume of the branes, with coupling strength $\gs N$. 
It is thus natural to
conjecture that the physics in the full three-brane 
metric
is dual to the worldvolume action for the 
D3-branes\footnote{A different approach to this same
problem was proposed 
in \cite{hashimoto}:
embedding the full D3-brane geometry in an asymptotically AdS 
background. The required geometry is simply that produced 
by two stacks of 
D3-branes a finite distance apart. 
By the standard AdS/CFT correspondence, 
this is dual to an $SU(N+K)$ SYM theory broken to $SU(N)\times SU(K)$.
It was found in \cite{hashimoto} that the information 
about the `full D3-brane' portion of the background is encoded in a 
very narrow energy range in the gauge theory, and is
consequently difficult to extract.
Absorption in the double-centered background was
considered recently
in \cite{costa}, and successfully compared with
the field theory result in \cite{costa2}.}
\cite{klebabs,ghkk,dealwis,gh}. 

Notice that, if we as usual 
wish to regard 
$\omega$ as being arbitrary, then
to comply with (\ref{restriction})
we must take $\ls\to 0$. To retain the non-conformal
information we are then 
forced to simultaneously send $\gs N\to\infty$
in such a way that $R\sim(\gs N)^{1/4}\ls$ remains fixed.
Even if we do not describe the limit this way, 
(\ref{restriction})
requires that the theory on the branes be
strongly coupled, $\lambda=g_{YM}^{2}N\gg 1$, in order for 
the curved picture supergravity background to be reliable.  
As emphasized in \cite{gh},
this requirement
implies that the worldvolume action
cannot be merely the Born-Infeld action. The latter 
arises from a disk-level string calculation, so
it does not
incorporate the effects of summing over worldsheets with
an arbitrary number of boundaries.

\subsection{Explicit conjectures}

To formulate an explicit duality conjecture, it is
thus necessary to determine the low-energy effective
action for a large number of
D-branes at strong 't~Hooft coupling--- undoubtedly a
daunting task. Fortunately, as explained
in \cite{gh,intri}, string-theoretic information highly 
constrains the possible form of the required action. First of all, 
the theory must reduce to $\cN=4$ SYM in the extreme infrared,
corresponding to the fact that the three-brane metric reduces to 
AdS for $r\to 0$. 
For
small but finite energy, the Lagrangian of the
dual theory can be expressed as
a deformation of the superconformal fixed point by irrelevant 
operators,
\be \label{irreldeform}
\cL=\cL_{SYM}+\sum_{d> 4}h_{d}R^{d-4}\cO_{d}~,
\ee
where $d$ denotes the dimension of the 
non-renormalizable operator $\cO_{d}$, and 
$h_{d}$ is a dimensionless coupling. 

The irrelevant operators $\{\cO_{d}\}$ 
ought to be compatible with the 
symmetries of the three-brane background: they must 
preserve sixteen 
supersymmetries (i.e.,  non-conformal $\cN=4$) and be 
invariant under the $SO(6)\sim SU(4)$ R-symmetry. 
The least irrelevant such operator is
\be \label{o8}
\cO_{8}= Q^{4}\bar{Q}^{4}\tr \Phi^{4}
       =\tr \left\{ F^{4}-{1\over 4}(F^{2})^{2}+\ldots \right\}~,
\ee
which
happens to be the leading correction to SYM obtained by expanding
the Born-Infeld action (see \cite{tseytlin} and references
therein).
As indicated schematically in (\ref{o8}), 
$\cO_{8}$ lies in a short
multiplet of the $\cN=4$ algebra: it
is a supersymmetric descendant
of the chiral primary operator $\tr\Phi^{4}$ (where the product 
of scalar fields is  understood to be symmetrized and traceless).  
$\cO_{8}$ is dual to a supergravity 
field $\pi$ which has mass-squared $m^{2}=32/R^{2}$ and describes 
deformations of the trace of the AdS$_{5}$ and $\bS^{5}$ metrics
\cite{krvn,ferrara2,lt}. 

Now, the AdS/CFT correspondence predicts that
for strong 't~Hooft coupling,
all operators in the gauge theory except those in short multiplets 
acquire large anomalous dimensions, $d\sim \lambda^{1/4}$. 
For $\lambda\gg 1$, then, the sum in (\ref{irreldeform}) is
effectively restricted to run only over
operators in short multiplets. All of the supergravity fields 
dual to such operators were tabulated in \cite{krvn}.  It is shown
there that the aforementioned field
$\pi$ is in fact the only scalar
$SO(6)$-singlet mode with positive mass-squared (i.e.,
dual to an irrelevant gauge theory operator).  Gubser and Hashimoto
\cite{gh}
were thus led to conjecture that, at least for $\gs\to 0$ and
$\gs N\to\infty$, physics on the curved three-brane
background is holographically encoded in the Lagrangian 
\be \label{conjecture}
  \cL=\cL_{SYM}+h_{4}R^{4}\cO_{8}~.
\ee
This was interpreted in  \cite{gh}
as a Wilsonian effective Lagrangian
with a cutoff of order $R$.

To test their conjecture, Gubser and Hashimoto 
computed the absorption probability $P$ for arbitrary
dilaton partial waves in the three-brane 
background\footnote{Three-brane absorption probabilities
for a broad class of massless modes with arbitrary energies
were determined in \cite{clvp}.},
exploiting the remarkable fact that the relevant equation
of motion has an exact solution in terms of associated
Mathieu functions \cite{gh} (see also
\cite{mmlz}). From $P$ it is possible to
deduce,
through an application of the optical theorem,
the two-point correlator of the gauge theory
operator dual to the dilaton partial wave under consideration. 
This exercise was carried out in \cite{gh}
for the dilaton s-wave,
employing the logic
explained in \cite{gk,gubser,magoo}. 
In Section 4 we will demonstrate that the form of the optical
theorem used in \cite{gh} is incomplete. Nevertheless,
the two-point 
function which satisfies the correct optical theorem (and which 
follows directly from the prescription we will
develop in the following 
section) has a form similar to the one
presented in \cite{gh}.
It was argued in \cite{gh}
that this form
could potentially be fully
explained in terms of the Lagrangian 
(\ref{conjecture}), although of course a 
perturbative calculation
would not be expected
to reproduce
the precise numerical coefficients
obtained at strong coupling from
the supergravity calculation. 
For the 
first correction to the conformal result,
this comparison was carried out already
in \cite{ghkk}.  A closely
related comparison can be found
in \cite{costa,costa2}.

The conjecture of Gubser and Hashimoto
was further analysed and 
considerably strengthened 
in subsequent work by Intriligator \cite{intri}. 
The last author arrived at (\ref{conjecture})
from a somewhat different perspective. 
His starting point is the assumption that there exists some
four-dimensional 
theory dual to the background (\ref{d3metric})
with an arbitrary $SO(6)$-symmetric harmonic function
\be \label{h}
H(r)=h+{R^{4}\over r^{4}}~.
\ee
Intriligator then
argues that the scaling properties of the metric
imply that $h$ should be interpreted as a coupling constant
which multiplies an operator
in the dual theory whose dimension is exactly eight
at all scales.
As stated before,
a renormalization group flow dual to the three-brane
background must preserve sixteen supersymmetries. 
{}From
a detailed analysis of flows with these many 
supersymmetries, 
the author of \cite{intri} 
concluded that 
along them 
the gauge
coupling constant does not run, and the dimensions of 
operators in 
short multiplets remain constant. 
In the case of immediate interest,
the former property is in line with the fact that the 
background dilaton
is constant, while the latter 
property implies that $\cO_{8}$ has dimension exactly eight along the 
entire flow. 
{}In this manner, 
Intriligator arrived at
the conclusion
that the 
background (\ref{d3metric}) with harmonic function (\ref{h}) is 
holographically dual
to the four-dimensional
gauge theory whose Lagrangian is exactly (\ref{conjecture})
along the entire flow, with $h_{4}\sim h$ \cite{intri}.
This duality statement
includes
the AdS$_{5}$/SYM$_{4}$ correspondence ($h=0$) as a particular case.
The case $h=1$ is of course the full
three-brane background, and all
other cases with $h>0$ are related to this by a rescaling of $r$ and 
$x^{\mu}$.
It was emphasized in \cite{intri} that the dimension-eight operator 
that
enters the duality could potentially be a linear combination
of the single-trace
operator (\ref{o8}) and the double-trace operator 
$\tr F^{2}\tr F^{2}+\ldots$, which has exactly the same quantum 
numbers.

While the information reviewed so far indicates that $\cO_{8}$ 
(possibly with some double-trace admixture)
is the only short-multiplet operator of relevance for the duality, 
it does not yet rule out that 
\emph{long}-multiplet $SU(4)$-symmetric 
operators preserving sixteen supersymmetries
could appear in the deformed
Lagrangian (\ref{irreldeform}).  If present, they
would be important for the duality conjecture
away from the strong-coupling regime. Intriligator argued
that this is in fact not possible, because
such operators would have to enter
the Lagrangian multiplied by a $g_{YM}$-dependent power of $h$, and 
this would lead to a non-trivial running of the gauge coupling 
constant, in contradiction with the fact that the 
dilaton in the supergravity background is constant  \cite{intri}.
Intriligator has thus conjectured that (\ref{conjecture}) is in fact 
the exact holographic dual of type IIB string theory on the background
(\ref{d3metric}) with harmonic function (\ref{h}), 
for any value of $\gs$ and $N$.
Notice that the duality has now been phrased in terms of the full 
string theory, as opposed to just supergravity: as explained before,
if $\gs N$ and $\omega R$ are arbitrary, then $\omega\ls$ is also 
arbitrary, implying that it is possible to excite higher
string modes. String excitations would
in fact be present on \emph{both} sides of the duality,
so it is not clear if a sensible meaning
can be ascribed to a duality
statement involving only the gauge theory modes\footnote{It seems
more sensible to speak of a closed string--open string duality, 
along the lines
of \cite{verlinde,kv} (see also \cite{pt,park}).}.

In a sense, the statement that (\ref{conjecture})
is dual to the full D3-brane geometry is a special case of the
AdS$_{5}$/SYM$_{4}$  correspondence. Given that
the operator $\cO_{8}$ is 
dual to the supergravity field $\pi$, the deformation of $\cL_{SYM}$ 
by $\cO_{8}$ should describe a
background which asymptotes to AdS space, with the mode $\pi$ excited. 

The standard situation  would be to consider 
$\cN=4$ SYM as a UV fixed point, and perturb away from it by adding 
\emph{relevant} operators $\cO_{d}$, $d<4$
(see \cite{magoo} and references 
therein). 
In that case the asymptotic $r\to\infty$ geometry
is AdS, with the appropriate supergravity field excited and having
a radial-dependence $r^{d-4}$.
The dimension of the operator and the mass of the dual field
are related through 
$d=2+\sqrt{4+m^{2}R^{2}}$ \cite{gkp,w}. 

On the contrary, for the deformation indicated in (\ref{conjecture})
one regards $\cN=4$ SYM as an IR fixed point.
The geometry is thus required to be asymptotically AdS as
$r\to 0$. Expanding the metric (\ref{d3metric}) with harmonic
function (\ref{h}), one finds that to linear order in $h$
\be \label{almostads}
ds^{2}\simeq ds_{AdS}^{2}+{h\over 2}\left({r\over R}\right)^{4}
\left[ -\left({r\over R}\right)^{2}
  (-dt^{2}+{d\mathbf{x}}^{2})+\left({R\over r}\right)^{2}dr^{2}
  +R^{2}d\Omega_{5}^{2} \right]~.
\ee
The form of the perturbation, and in particular
its dependence  on $r^{4}$ (relative
to AdS), is consistent with a deformation
associated  with the field $\pi$, whose mass satisfies
$m^{2}R^{2}=32$ \cite{cm}.
As $r$ increases,
the metric perturbation indicated in (\ref{almostads})
grows large, so it becomes necessary to 
solve the full non-linear supergravity equations. The complete 
solution is of course the three-brane background, which 
differs drastically from AdS at large $r$.

On the gauge theory side, we are attempting to define the theory 
starting from the IR fixed point and 
following the RG flow in the reverse direction. 
The presence of a
non-renormalizable interaction would ordinarily
point to the need for a new definition of the theory in the UV.  
On this issue, Intriligator espoused the view that (\ref{conjecture})
describes the theory at all scales, with the understanding that the 
coefficients of all other
irrelevant operators are fine-tuned to zero \cite{intri}.

We do not necessarily subscribe to this view. 
In particular, we should
stress that, if D-branes are invoked
to motivate the duality, then to us it seems inevitable that
excited open string modes
enter the duality at super-stringy energies, $\omega>\ms$. Such
energies  can certainly be reached if both $R$ and
$\gs N$ are finite. On the other hand, Intriligator's line of
argument does not really
equate the dual theory with 
the  D3-brane worldvolume theory:
the duality is asserted to
hold for arbitrary $\gs$ and $N$, whereas
we know that (\ref{conjecture}) certainly 
does not summarize the low-energy dynamics 
of a small number of weakly coupled branes, $\gs\ll 1$, $N\sim 1$.   
So whether or not it is correct, this strong form of the duality
conjecture cannot be said to rest directly on D-brane intuition. 
Instead, it is based on
the (by now fairly standard) assumption
that \emph{any} theory
of gravity can be described holographically through
a lower-dimensional non-gravitational theory \cite{thooft,susskind}.

\subsection{Our approach}

Given the immense difficulties encountered
in attempting a direct analysis of
the candidate dual theory, 
we choose to follow an indirect route.
We assume that there exists \emph{some} theory
which is the holographic image of 
physics on the three-brane background.
We will henceforth refer
to this theory as the `holographic dual'.
Based on this existence assumption, in the following sections
we will develop calculational tools
that allow us to compute quantities in the holographic dual 
in terms of supergravity.  The spirit of our approach is
similar to that of \cite{abks,ms,mr}, in that we use a 
conjectural duality to extract information about the 
dual theory. 
Since our calculations are based 
entirely on the supergravity side of the duality,    
our results contain information about
whatever theory turns out to be the holographic dual of the D3-brane 
background, even if it is not of the form (\ref{conjecture})
along the entire RG flow. 
Notice in particular that 
the fact that the supergravity
solution is smooth for all $r$
presumably indicates
that the dual theory is
sensible at all energy scales.
We regard
our work as a step towards a more 
precise specification of this dual theory.

\section{Holographic Two-Point Function from Supergravity} 
\label{2ptsec}

If there exists
a duality relating supergravity in a D3-brane
background to some four-dimensional theory, 
the duality mapping ought to work in 
both directions. By definition, the lower-dimensional theory
should holograph supergravity in the ten-dimensional background.
Conversely, the physics of the holographic theory should
be encoded in the bulk.
In particular,
a prescription should exist for computing correlation 
functions of the dual theory in terms of
the curved D3 spacetime. In this section we will construct such a 
prescription.
For definiteness, we focus on the two-point function
of the operator dual to the dilaton
s-wave. The extension to higher partial waves is
straightforward; the main steps are described in Appendix B,
and the final result is quoted at the end of this section.
We will generalize the recipe to 
higher correlators in Section \ref{nptsec}.

Before we proceed, we should take a moment to indicate why
we have found it necessary to develop a new prescription. 
In the AdS/CFT case
it has been argued \cite{bgl,gidd}
that the GKPW recipe
\cite{gkp, w} essentially equates correlation 
functions in the CFT with (generalized) AdS scattering 
amplitudes. It is thus natural to guess that the GKPW 
prescription can be carried over
to our setting by simply replacing the AdS 
bulk-to-boundary propagator with the corresponding solution in the 
full D3 background. Unfortunately, this guess
yields a two-point function which
fails to reproduce the correct absorption probability, 
and suffers from severe problems at high energies\footnote{ 
We are referring here to the appearance
of UV divergences with non-standard momentum-dependence.
Extrapolation of the GKPW recipe led to similar high-energy
problems
in \cite{ms,mr}, which were dealt with using \emph{ad hoc} 
momentum-dependent multiplicative renormalizations. 
The need for this peculiar procedure
was taken as an indication
of the non-local nature of the corresponding
dual theories.}.  
In this section, following a
more physical approach, we 
will be arrive at a prescription  which
gives a two-point function 
in accord with the absorption
results (see Section \ref{abssec}), and has a built-in 
subtraction and amputation
procedure which 
automatically ensures a well-defined UV limit.

\subsection{The prescription}

Consider a dilaton propagating in the presence of $N$ 
D3-branes. Let us denote the
corresponding propagator by  $G(r,r')$ (for the time
being we focus only on the $r$-dependence;
the other nine directions are left implicit). 
Viewing the branes as a supergravity solution,
this propagator is obtained to lowest order in \gs\
by solving the linearized
equation of motion for the dilaton in the curved ten-dimensional
background. Higher-order corrections, involving supergravity 
interactions, are suppressed in the $\gs\to 0$
limit. In taking this limit,
we keep the geometry fixed, i.e., we 
hold $R^{4}\propto \gs N$ constant.

On the other hand, we can view the D-branes as (3+1)-dimensional
objects with intrinsic
dynamics, embedded in
a flat (9+1)-dimensional ambient spacetime. In
the obvious coordinate system,
they are localized (along six directions) at $r=0$. 
The $l$th partial wave of the
\emph{canonically normalized} dilaton $\phi$ couples with
unit strength to an operator in the worldvolume
theory which will be denoted
by $\cO_{\phi}$ (the angular-momentum labels of these
operators will be left implicit). The leading low-energy
terms of these operators
can be found in \cite{ktvr}.
The dilaton s-wave, in particular, couples to the
operator
\be \label{o}
\Ophi= -{\sqrt{2 \pi}\over 4} \tr \left\{
   (\sqrt{2}\pi R^{2})^{2}F^{2} + 
   (\sqrt{2}\pi R^{2})^{4}\left( F^{4}-{1\over 4}(F^{2})^{2}\right)
   +\ldots
\right\}~,
\ee
where the `$\ldots$' represent scalar and fermion dimension-four
operators
present already in the conformal limit \cite{ktvr}, as well as 
additional operators
of dimension eight and possibly higher
which are corrections away 
from this limit \cite{ghkk}. 
We remind the reader that the field strength $F$ in (\ref{o}) is 
't~Hooft-normalized, i.e., the combination
which appears in the worldvolume action is
$N\tr F^{2}$.  

\begin{figure}[htb]
\centerline{\epsfxsize=\textwidth\epsfbox{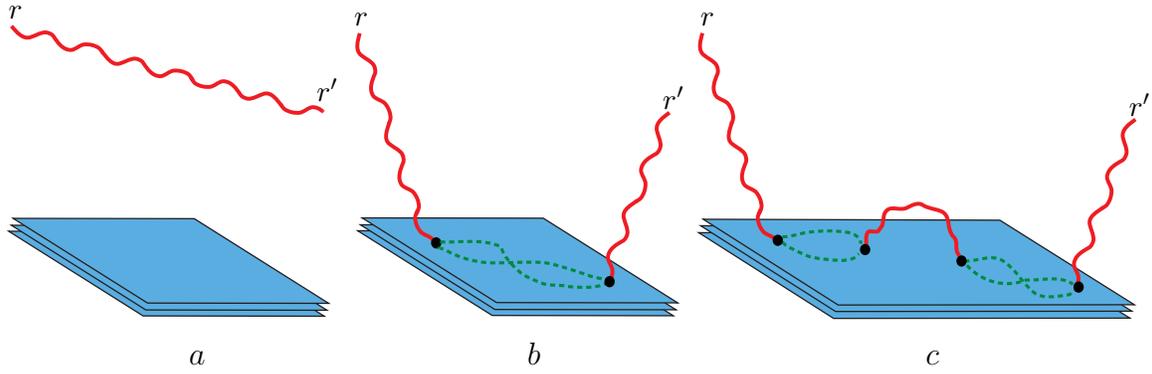}}
 \begin{picture}(0,0)
   \put(25,0){$a$}
   \put(70,0){$b$}
   \put(123,0){$c$}
   \put(1,46){\small $r$}
   \put(47,45){\small $r$}
   \put(93,45){\small $r$}
   \put(42,35){\small $r'$}
   \put(88,34){\small $r'$}
   \put(150,33){\small $r'$}
 \end{picture}
\caption{\small Propagation of a flat space dilaton in the presence of 
D3-branes. Starting from point $r$, the dilaton can reach $r'$ a) 
directly, or b,c) indirectly, after having interacted with the branes.
Black dots denote insertions of the operator $\cO_{\phi}$, and 
dotted lines represent worldvolume processes.}
\end{figure}

A dilaton can propagate from $r$ to $r'$ either directly 
(Fig.~1a), or indirectly, after having interacted
with the branes  
(e.g., as depicted in Figs.~1b,c)
by means of the coupling
$\int d^{4}x\,\phi\Ophi$.
This results in a series of
contributions to the propagator which are
expressed diagrammatically in Fig.~2, 
where $C_{n}$ denotes the 
`pure worldvolume' connected $n$-point correlator of $\Ophi$
(i.e., the correlator computed exclusively with the brane action
$S_{b}$ described in Section 2). 

\begin{figure}[htb]
\centerline{\epsfxsize=\textwidth\epsfbox{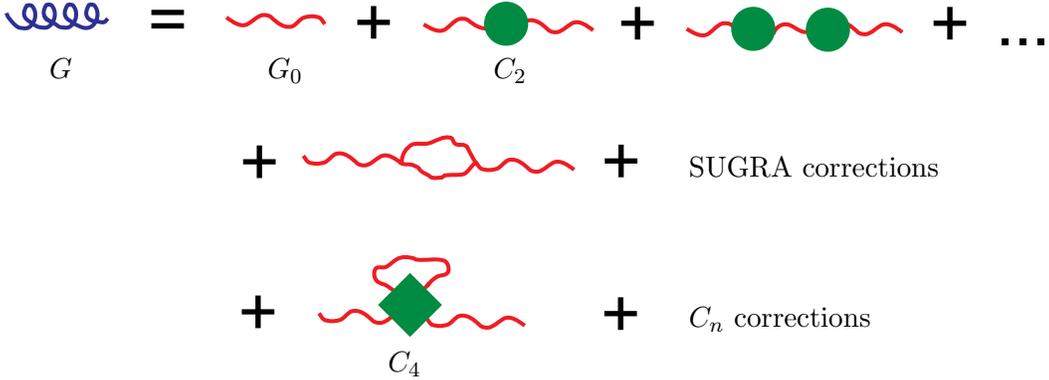}}
 \begin{picture}(0,0)
  \put(15,53){\small $G$}
  \put(44,53){\small $G_{0}$}
  \put(74,53){\small $C_{2}$}
  \put(100,40){\small SUGRA corrections}
  \put(60,14){\small $C_{4}$}
  \put(100,20){\small $C_{n}$ corrections}
 \end{picture}
 \vspace*{-1cm}
\caption{\small Diagrammatic expansion for the two-point dilaton
correlator. The full propagator $G$
is written out in terms
of the flat space propagator $G_{0}$, 
flat space supergravity vertices, and `pure-brane' $n$-point
correlators $C_{n}$. The first three diagrams on the 
right-hand side correspond to processes of the type shown
in Figs.~1a, 1b and 1c, respectively. See text for discussion.}
\end{figure}

If we take the `decoupling' limit $\gs\to 0$ with 
the 't~Hooft coupling $\lambda\propto \gs N$ fixed, 
the expansion in Fig.~2 
simplifies drastically. Diagrams involving supergravity 
vertices evidently vanish. In addition, almost all
diagrams involving brane correlation functions drop out. The 
$\gs$-dependence of the correlators is known in the conformal limit,
from the standard AdS/CFT correspondence. According
to the GKPW recipe \cite{gkp,w}, boundary theory correlators are 
obtained from bulk AdS diagrams with $n$ external dilaton legs which 
terminate at the boundary. $C_{2}$ is consequently $\gs$-independent, 
since the relevant graph is just the propagator. The graphs for
(connected) higher-order correlators feature supergravity vertices, 
and as a result, (the leading large-$N$ contribution to) 
$C_{n}$ is proportional to $\gs^{n-2}$. 
From the point of view of the field 
theory, this corresponds to the 
large-$N$ factorization of correlation functions.
It is thus clear 
that in the $\gs\to 0$ limit, the expansion in Fig.~2
collapses to
\bea \label{propexp}
G(r,r')&=& G_{0}(r,r')+G_{0}(r,0)C_{2}G_{0}(0,r')
          +G_{0}(r,0)C_{2}G_{0}(0,0)C_{2}G_{0}(0,r')+\ldots 
          \nonumber \\
{}&=& G_{0}(r,r')+G_{0}(r,0)\Delta_{2}G_{0}(0,r'),
\eea
where $G_{0}$ is the flat-space dilaton propagator, and in the second 
line we have denoted by $\Delta_{2}$ 
the sum of the indicated series.

It might seem surprising that even at vanishing $\gs$ the
branes are capable of emitting and reabsorbing dilatons.  
{}From the 
string theory perspective, diagrams with additional
closed strings usually have 
extra handles, and are thus suppressed as $\gs\to 0$. However,
what looks like an extra handle in going from, e.g.,
Fig.~1b to Fig.~1c, 
in fact sews together two surfaces which would otherwise be disjoint.
The net effect is to add some number of boundaries
to the original surface.
In this manner, 
each successive 
term in the first line of (\ref{propexp}) originates from  a
string diagram with additional
worldsheet boundaries, which contribute additional 
powers of $\gs N$, not of $\gs$. 
Since we work in the regime of strong 't~Hooft 
coupling, the entire series in (\ref{propexp}) must indeed be kept. 
Notice this means that there is no way to
disentangle processes like those shown in Fig.~1c from the 
`purely worldvolume' graphs that are contained in $C_{2}$. String
theory thus dictates that it is 
$\Delta_{2}$, and \emph{not} $C_{2}$, which must be regarded as the 
two-point correlator of $\Ophi$,
\be \label{delta2def}
\Delta_{2}=\brack{\Ophi\,\Ophi}~,
\ee
in the effective theory
summarizing the dynamics of the `decoupled' brane system.
This is the theory which can be 
expected to holograph the physics of the curved D3 
background.

If we identify $G(r,r')$ in (\ref{propexp}) with the curved space
dilaton propagator, then the equality can {\it a priori \/} 
only be expected
to hold in the limit $r,r'\to\infty$, because it is
only far away from the branes
that one can meaningfully compare $G$ with the flat
space propagator $G_{0}$. 
The essential
point here is that, if we 
took (\ref{propexp}) as it stands as
our definition of $\Delta_{2}$, then we would
expect $\Delta_{2}$ to depend on $r,r'$, complicating its 
interpretation as a correlator in a four-dimensional
theory.
Our main goal in the remainder of this
section and the next will be to show that, on the contrary,
\be \label{delta2}
\Delta_{2}= \lim_{r,r'\to\infty}\frac{G(r,r')-G_{0}(r,r')}
           {G_{0}(r,0)G_{0}(0,r')}
\ee
is a well-defined quantity which can be rightfully 
interpreted as the desired
four-dimensional holographic
correlator.\footnote{Once this is established, it 
becomes natural to speculate, based on the insight gained 
from AdS/CFT \cite{susswi,pp},
that (\ref{propexp}) holds also for finite 
$r,r'$, with $\Delta_{2}$ interpreted as the two-point function
with a UV cutoff (or more precisely, with $r$ and $r'$ 
indicating a `smearing' of the
two insertions of $\Ophi$).}

\subsection{Calculation of the correlator}

To examine the limit $r,r'\to\infty$, we first need to discuss the 
propagators in more detail.  In order for the essential points
to be more easily appreciated, we will first
carry out the calculations at a general level, 
postponing explicit evaluations
to the next subsection. 
The propagators $G$ and $G_{0}$ are defined as solutions
to the ten-dimensional dilaton equation of motion
\be \label{10dpropeqn}
\p_{M}\left[\sqrt{-g}\,g^{MN}\p_{N}G\right]=\delta^{(10)}(X-X')
\ee
in the respective backgrounds.
Upon projecting onto a plane wave $\exp(i k_{\mu}x^{\mu})$
for the directions parallel to the 
three-branes, and onto the constant mode on $\bS^{5}$, 
one is left with an equation for
the radial propagator $G(k^{\mu};r,r')$ of the form
\be \label{propeqn}
\left\{\p_{r}^{2}+{\p_{r}f\over f(r)}\p_{r}+{V_{k}(r)}
  \right\}G= {\delta(r-r')\over \pi^{3} f(r)},
\ee
where $f(r)=\sqrt{-g}g^{rr}$.
Since we regard $G(k;r,r')$ as the field created at $r$ by a source 
at $r'$, it must satisfy boundary conditions such that (for 
$\omega\equiv k^{0}>0$) the associated
flux moves \emph{away} from the source.

At $r\not= r'$, (\ref{propeqn}) is just the homogeneous equation for radial
motion of the dilaton. Denote the solution with the required behaviour 
at $r\to 0$ (and consequently at all $r<r'$) by $\phi_{1}$.
Similarly, let $\phi_{2}$ be the solution of
the homogeneous equation obeying
the appropriate boundary condition at $r\to\infty$.
Finally, take 
$\phi_{3}$ to be the solution which is linearly independent from
$\phi_{2}$ and 
satisfies the `opposite' boundary condition at $r\to\infty$.
Of course, the solutions
$\{\phi_{1},\phi_{2},\phi_{3}\}$ are not independent,
and we can write 
\be \label{phi1}
\phi_{1}=A\phi_{3}+B\phi_{2}~, 
\ee
with $A,B$  some overlap
coefficients. 
It is clear that the $r$-dependence of the propagator
can be expressed in the form
\be 
G(r,r')=\left\{ {{\alpha(r') \phi_{1}(r) \quad \mbox{if}\quad r<r'}\atop
                {\beta(r') \phi_{2}(r) \quad \mbox{if}\quad r>r'}} 
                \right.
\ee
for some $\alpha,\beta$. These functions of $r'$ can be determined 
by demanding that the propagator be continuous at $r=r'$, and 
its first derivative have a discontinuity at $r=r'$ which
yields the delta-function
in (\ref{propeqn}), with unit coefficient. The end result is
\be \label{prop}
G(r,r')={1\over w_{32}A}\phi_{2}(r_{>})\phi_{1}(r_{<})
       ={1\over w_{32}}\phi_{2}(r_{>})\left[\phi_{3}(r_{<})
       +{B\over A}\phi_{2}(r_{<})\right],
\ee
where $r_{<}$ ($r_{>}$) is the smaller (larger) of $r$ and $r'$, 
$w_{32}$ is the constant appearing in the Wronskian
$W_{32}(r)=\phi_{3}\p_{r}\phi_{2}-\phi_{2}\p_{r}\phi_{3}=w_{32}/f(r)$,
and $A,B$ are the overlap coefficients defined in (\ref{phi1}).

Now we have enough information
to explore the nature of the limit in (\ref{delta2}). 
By definition, the flat space
solution $\phi_{02}$ ($\phi_{03}$) 
must asymptote to a purely outgoing (ingoing) wave
as $r\to\infty$: 
$\phi_{02}(r)\to r^{-5/2}\exp[i(qr+\theta_{0})]$.  This must be 
true as well for the corresponding solution in the 
asymptotically flat D3 
background, with the same $q$  
but a different `phase shift', $\theta$. 
Finally, the flat space 
solution $\phi_{01}$ 
must be regular at the origin, which implies that
$A_{0}=B_{0}$, and 
$G_{0}(r,0)\to r^{-5/2}\exp[i(qr+\theta_{0})]$. 
Using all of this 
in (\ref{delta2}) one is left with
\be \label{bo}
\Delta_{2}= {1\over w_{32}}\left[{w_{012}\over \phi_{01}(0)}
 \right]^{2}
\Bigg[ {B\over A}e^{2i(\theta-\theta_{0})}-1
     \Bigg]~,
\ee
where $A,B$ are the curved space overlap
coefficients and $w_{012}/f(r)$ is the Wronskian of the indicated
flat-space solutions.
As advertised, all $r,r'$-dependence cancels out, and the limit is 
well-defined. This result is as expected from our 
derivation of (\ref{delta2}), and serves as a first consistency check on 
our approach.

\subsection{Explicit evaluation}

Having explained the essential points, let us now proceed to the 
explicit determination of the dilaton propagators. Using the
metric (\ref{d3metric}) in (\ref{10dpropeqn}), the curved space 
propagator $G(k^{\mu};r,r')$ can easily be seen to
satisfy (\ref{propeqn}) with $f(r)=r^{5}$ and
$V_{k}(r)=q^{2}H(r)$, 
where we have defined
$q^{2}=\omega^{2}-\vec{k}^{2}$. 
As explained in \cite{gh} (see 
also \cite{mmlz}),
the corresponding homogeneous equation can be related to 
Mathieu's equation, and the solutions of interest to us are found to
be
\bea \label{mathieu}
\phi_{1}(r)&=&r^{-2}H^{(1)}(\nu,-\ln(r/R))~, \nonumber\\
\phi_{2}(r)&=&r^{-2}H^{(1)}(\nu,+\ln(r/R))~, \nonumber\\
\phi_{3}(r)&=&r^{-2}H^{(2)}(\nu,+\ln(r/R))~. 
\eea
Unless otherwise noted,
we adopt the notation of \cite{gh}: 
$H^{(1,2)}(\nu,z)$ are associated Mathieu functions of the third and 
fourth kind, respectively, and 
$\nu$ is the `Floquet exponent' (an $l$-dependent function of $qR$)
defined in \cite{gh,mmlz}. 
In line with the previous discussion,
the following boundary conditions have been enforced:
$\phi_{1}$ is
purely ingoing at the horizon $r=0$, while $\phi_{2}$ ($\phi_{3}$) 
is purely 
outgoing (ingoing) at $r\to\infty$. 
The Wronskian of $\phi_{2}$ and 
$\phi_{3}$ works out to
$w_{32}=4i/\pi$. 
For future use, we note
that with these boundary conditions, (\ref{phi1}) implies that 
the absorption probability is given by
\be \label{pabs}
P_{\mathrm{abs}}= 1-\left| {B \over A} \right|^{2}~.
\ee

{}From Eq.~(18) in \cite{gh} we can read 
off the superposition coefficients
\be \label{AB}
A=\frac{\chi-{1 \over \eta^{2}\chi}}{{\eta-{1\over\eta}}}, \qquad 
B=\frac{\chi-{1\over\chi}}{{\eta-{1\over\eta}}},
\ee
where $\eta=\exp(i\pi\nu)$ and $\chi=\varphi(-\nu/2)/\varphi(\nu/2)$, with
$\varphi(\pm\nu/2)$ 
(not to be confused with the radial solutions 
$\phi_{i}(r)$) two of the coefficients involved in the definition of 
Mathieu functions \cite{gh,mmlz}.  Both $\eta$ and $\chi$ are 
functions of $qR$, which we will
characterize further in Section \ref{abssec} and Appendix A. 
Using all of this
in (\ref{prop}) we obtain 
\be \label{d3prop}
G(k^{\mu};r,r')=-{i\pi \over 4  r^{2}r'^{2}}
   H^{(1)}(\nu,\ln ({r_{>}\over R})) 
   \left\{H^{(2)}(\nu,\ln ({r_{<}\over R}) ) +
   \frac{\chi-{1\over\chi}}{\chi-{1\over \eta^{2}\chi}}
   H^{(1)}(\nu,\ln ({r_{<}\over R}) )
   \right\}.
\ee
As $r\to\infty$, the Mathieu functions asymptote to Hankel functions, 
and 
\be \label{d3propasym}
G(k^{\mu};r,r')\to -{i \over 2 q}\left({1\over r r'}\right)^{5/2}
   e^{i(qr_{>}+\theta)}\left\{e^{-i(qr_{<}+\theta)} +
   \frac{\chi-{1\over\chi}}{\chi-{1\over \eta^{2}\chi}}
   e^{i(qr_{<}+\theta)}
   \right\},
\ee
with $\theta=-\pi(2\nu+1)/4$.

The derivation of the flat space propagator proceeds in the same 
steps. The solutions to the homogeneous version of (\ref{propeqn}) are now
just Bessel and Hankel functions,
\bea \label{bessel}
\phi_{01}(r)&=&r^{-2}J_{2}(qr)~, \nonumber\\
\phi_{02}(r)&=&r^{-2}H_{2}^{(1)}(qr)~, \nonumber\\
\phi_{03}(r)&=&r^{-2}H_{2}^{(2)}(qr)~. 
\eea
The appropriate boundary condition at the origin
$r=0$ is now simply regularity
of the solution, which picks out $J_{2}$ as the correct solution. The 
propagator is found to be
\be \label{flatprop}
G_{0}(k^{\mu};r,r')=-{i\pi\over 4 r^{2}r'^{2}}
   H^{(1)}_{2}(qr_{>}) \left\{H^{(2)}_{2}(qr_{<}) 
   + H^{(1)}_{2}(qr_{<})
   \right\},
\ee
so in the limit $r,r'\to\infty$ it becomes
\be \label{flatpropasym}
G_{0}(k^{\mu};r,r')\to -{i\over 2 q}\left({1\over r r'}\right)^{5/2}
   e^{i(qr_{>}+\theta_{0})}\left\{e^{-i(qr_{<}+\theta_{0})} +
   e^{i(qr_{<}+\theta_{0})}
   \right\}~,
\ee
with $\theta_{0}=-5\pi/4$.
Additionally, as $r\to\infty$ one finds
\be \label{flatprop0infty}
G_{0}(k^{\mu};r,0)\to {i\sqrt{2\pi}\over 16}
  \left({q^{3/2}\over r^{5/2} } \right) e^{i(qr+\theta_{0})}.
\ee

Inserting Eqs.~(\ref{d3propasym}), (\ref{flatpropasym}) 
and (\ref{flatprop0infty}) into (\ref{delta2}) we finally arrive at
an explicit expression for the two-point function of $\Ophi$ in
the holographic theory,
\be \label{delta2exp}
\Delta_{2}(q^{2})={64\pi^{2} \over  q^{4}} i \left[ 
    \frac{\chi-{1\over\chi}}{\eta\chi-{1\over \eta\chi}} -1\right]~.
\ee

The preceding calculation has focused
on the $\bS^{5}$-symmetric mode of the dilaton, but the prescription
we have derived generalizes 
to arbitrary supergravity fields in a straightforward manner.
In particular, 
one can easily determine the two-point functions 
of the operators dual to all dilaton partial waves. 
The general result is 
\be \label{delta2lexp}
\Delta^{(l)}_{2}(q^{2})=2^{5+l}\pi^{2}(l+1)(l+2) q^{-2l-4} i \left[ 
    (-)^l\frac{\chi_l-{1\over\chi_l}}
    {\eta_l\chi_l-{1\over \eta_l\chi_l}} -1\right]~.
\ee
The main steps of the 
calculation are given in Appendix B.
The analysis of the above result
will be the subject of the next section.

\section{Low Energy Limit and Absorption Probability} \label{abssec}

In order to understand the properties of the correlator
(\ref{delta2lexp})
it is convenient to introduce the following notation:
\be
s=(qR)^2,\ \ \chi_l = e^{i\mu_l(s^2) \ln s + i \alpha_l(s^2)},\ \
\nu_l=l+2+i\mu_l(s^2).
\ee
It can be inferred from the results of \cite{gh,mmlz} (summarized in
Appendix A)
that $\alpha_l(s^2)$ and $\mu_l(s^2)$ are
analytic functions  (in a neighborhood of $s=0$)
which, for real $s$, are real when $l=0$ and purely imaginary for $l>0$.
In this notation,
Eq.~(\ref{delta2exp}) (which is valid for the case $l=0$) can be rewritten
as:
\be \label{cotgformula}
\Delta_{2}(s) = -{64\pi^2 R^{4}\over s^{2}}
    {\mathrm{cot}}\left[ \mu_{0} \ln(-s) + \alpha_{0}\right]
\sinh(\pi\mu_{0})
    +{64\pi^2 R^{4}\over s^{2}} i \left(\cosh(\pi\mu_{0})-1\right).
\ee
The second term contains only (even) integer powers of $s$ in an expansion
around $s=0$ and therefore corresponds to contact terms.
We can drop it in a comparison
to field theory. The first term has a cut for positive real $s$ where the
imaginary
part changes sign, being negative above the real axis and positive below.
In addition,
it becomes real for negative real $s$. 
It is clear then that the first term has the
analytic properties expected from a field theory propagator. 
On the other hand, 
even if the second term is dropped 
on the grounds that it is analytic, it is somewhat unsettling
that it is imaginary for real values of $s$. We will return to this 
point below.

Using the above notation,
it is equally easy to write down the two-point function for a generic value
of angular momentum $l$ (see Appendix B for the derivation):
\be \label{cotgformula_l}
\Delta^{(l)}_{2}(s)=- 2^{l+5}(l+1)(l+2) {\pi^{2}R^{4}\over s^{2}}
    {\mathrm{cot}}\left[ \mu_l \ln(-s) + \alpha_l\right] \sinh(\pi\mu_l) +
    \mathrm{analytic}.
\ee
The $l$-dependent factor in front arises
from the normalization factor of the spherical harmonics.
Incidentally, notice that  our method yields a finite result,
 in contrast with the
AdS calculation \cite{gkp,w,magoo}, which produces
a divergent result that needs to be renormalized.

\subsection{Low energy limit}

Using the results of \cite{gh,mmlz}, which we summarize in Appendix A, the
small $s$ behaviour of $\Delta^{(l)}_2$ follows as
\be \label{AdSlimit}
\Delta^{(l)}_{2}(s) = {\mathrm{analytic}}
            -\frac{\pi^3}{2^{3l+1}l!(l+1)!^3(l+2)} R^{4} s^{l+2} \ln(-s)
     + {\mathrm{higher\ order\  in\ }}s.
\ee
The low-energy behaviour of this propagator is fixed by conformal
invariance up to a normalization constant.
 It was computed directly in the SYM theory in \cite{ktvr}.
 To compare with that result one has
to multiply $\Delta^{(l)}_2$ by $s^l C_p^9 C_p^9$ (using the notation of
\cite{ktvr}),
from their definition of the operators coupling to the
dilaton,
and also
divide by a factor $2\kappa^2$,
from the difference between the standard and canonical
dilaton normalization. After
including
all these factors we obtain the result of \cite{ktvr} up to an overall
factor of two. Notice that one should not expect exact agreement,
since as explained in Section \ref{2ptsec},
the propagator $\Delta_{2}$ incorporates not only
worldvolume processes but also the coupling with 
the flat space dilaton. 
It is related to the `pure brane' two-point function $C_{2}$ through
Eq.~(\ref{propexp}).

 For the case $l=0$ the result including more than just the leading 
 term is 
\bea
\Delta_2(s) &=&-\frac{\pi^3}{4}  R^{4}s^2\ln(-s)
 \left[ 1-\frac{1}{24} s^2 \ln(-s) 
+\frac{7}{72} s^2  +\frac{17}{6912} s^4 (\ln(-s))^2 \right. \\
{}&{}& 
 - \frac{161}{18432} s^4 \ln(-s)+\frac{5\pi^2}{13824}  s^4
 \left.
 +\frac{5561}{663552} s^4 + \cdots \right]+ \mathrm{analytic}~. \nonumber
\eea
The first five terms
agree with those that can be deduced from the results of 
\cite{gh}, up to the same
overall factor of two (the first two terms are implicit already 
in \cite{ghkk}). 
The sixth term (of order $s^{6}\ln(-s)$) disagrees with \cite{gh};
it is the first discrepancy due to the fact that, as explained in 
the following subsection, the analysis of \cite{gh} does not employ 
the correct form of the optical theorem. 

It would be interesting to examine the high-energy behaviour of the
propagator. This, however, would require a better understanding 
of the role played by the second term in (\ref{cotgformula}) and 
(\ref{cotgformula_l}). We have already pointed out that this term is 
peculiar, and we will have more to say about it below. 
Let us just note that, as will 
be discussed in the next subsection, $\Delta_{2}^{(l)}$ 
is related to the absorption
probability $P_{\mathrm{abs}}$ 
through Eq.~(\ref{Pabs}). Since $P_{\mathrm{abs}}\to 1$
as $s\to\infty$, we seem to conclude that the UV behaviour
of the propagator is
$\Delta_{2}^{(l)}(s)\sim s^{-l-2}$.
Extracting this result directly from (\ref{delta2lexp}) is a 
delicate matter. Consider for instance the case $l=0$.
It has been pointed out in \cite{mr} 
that a WKB analysis 
shows that the first term of (\ref{delta2exp}) is
a decaying exponential at 
high energies. The expected $q^{-4}$ behaviour thus appears to 
arise entirely from the second term, which has the peculiar 
property of being purely imaginary. 

\subsection{Absorption probability}

{}From the field theory point of view the propagator computed in the
previous
section can be related to the absorption cross-section using unitarity.
Here we proceed to compute such relation and then, as a consistency check,
verify that it is satisfied by our propagator.
 Writing the S-matrix as $S=1+i \cT$, the well known identity
\be \label{opth-gen}
-2 \Ima \langle i|{\cT} |i\rangle  = \langle i |
\cT^\dagger \cT |i\rangle
\ee
follows.
 The state $|i\rangle$ will be taken to be an incident dilaton moving
with momentum $k^{\mu}$ parallel to the brane and with transverse momentum
$\vq$. Let $x_{0\ldots 4}$ and $y_{5\ldots 9}$ denote coordinates
parallel and perpendicular to the brane, respectively. 
The coordinate $y_9$ is chosen to be parallel to
the incident dilaton, i.e, $\vq=q\hy_{9}$.
 The S-matrix follows from the interaction
\be \label{opth-int}
 S_{\mathrm{int}} = \int d^4x
  \left.{\sum_{{{i_{1}} \ldots {i_{l}}=5}}^9}
  \partial_{{i_{1}} \ldots {i_{l}}} \phi(x,y) \right|_{\vy=0}
  C^{i_1\ldots i_l}_{l\vm} \cO^{l\vm}.
\ee
where the coefficients $C^{i_1\ldots i_l}_{l\vm}$ are defined in
Appendix B.
All supergravity interactions are suppressed by powers of $g_s$ and
so discarded in the limit $g_s\rightarrow 0$.
The field $\phi(x,y_\perp)$ is expanded as
\be
\phi(x,y_\perp) = \int \frac{d^3kd^6q}{(2\pi)^9 2\omega_{k,q}}
 \left( e^{-ik\cdot x-iq\cdot y} a_{k,q} 
 + e^{ik\cdot x+iq\cdot y} a^\dagger_{k,q} \right)
\ee
where
the frequency $\omega_{k,q}\equiv k^{0}=\sqrt{\vk^2+\vq^2}$, 
and the operators $a_{k,q}$
and
$a^{\dagger}_{k,q}$ satisfy
\be
{}[a_{k',q'},a^{\dagger}_{k,q}] = (2\pi)^9 2\omega_{k,q} \delta^{(4)}(k-k')
                                  \delta^{(6)}(q-q') .
\ee
With these conventions and using the fact
that the only interaction vertex where
$\phi$ appears is the one in Eq.~(\ref{opth-int}) we get
\bea
\langle 1_{k,q} | \cT | 1_{k,q} \rangle &=& - VT (-)^l
 \int d^4x e^{ikx}\langle 0| \hat{T} \cO^{l\vm}(x)\cO^{l\vm}(0)|0\rangle
C^{i_1\ldots i_l}_{l\vm}C^{j_1\ldots j_l}_{l\vm}
q_{i_1}\ldots q_{j_l} \nonumber\\
&=& VT (-)^l i\Delta^{(l)}_2(k) q^{2l} \sum_{\vm}
C^{9\ldots 9}_{l\vm} C^{9\ldots 9}_{l\vm},
\eea
where $V$,$T$ are normalization volume and time (which cancel in the final
result) and in the second equality we used the fact that the incident
particle propagates along $y_9$. Note also that the $\cO^{l\vm}$ propagator
is computed including the vertex (\ref{opth-int}).

 To compute $\langle 1_{k,q} |\cT^{\dagger} \cT| 1_{k,q}\rangle $ one has
to insert a complete set of states between $\cT^\dagger$ and $\cT$. 
With the only
interaction being
(\ref{opth-int}), the possible processes are elastic scattering
and absorption. Elastic scattering is described by
inserting
one-dilaton states. The calculation
is similar to the previous one:
\be
\begin{array}{rl}
\int {{d^3k'd^6q'}\over{(2\pi)^9 2\omega_{k',q'}}} &
\langle 1_{k,q} | \cT^\dagger | 1_{k',q'} \rangle
\langle 1_{k',q'} | \cT | 1_{k,q} \rangle \ =  \\ &\\
=& \frac{VT}{2(2\pi)^5} q^{4l+4}
\sum_{\vm \vm'}
C^{9\ldots 9}_{l\vm} C^{9\ldots 9}_{l\vm'} |\Delta^{(l)}_2|^2
\int d\Omega_{q'}
 C^{i_1\ldots i_l}_{l\vm}C^{j_1\ldots j_l}_{l\vm'}
\hat{q}'_{i_1}\ldots \hat{q}'_{j_l} \\ &\\
=& \frac{VT}{2(2\pi)^5}q^{4l+4} \cNl \sum_{\vm}
C^{9\ldots 9}_{l\vm} C^{9\ldots 9}_{l\vm} |\Delta^{(l)}_2|^2.
\end{array}
\ee
Here $\hat{q'}$ is a unit vector indicating the direction of the outgoing
particle ($|\vec{q}^{\,\prime}|$ 
is fixed by energy conservation) and $\cNl$ is the
result of the angular integration, which can be found in Appendix B.

 The other contribution is from inelastic scattering, which in our case
 is simply proportional to the absorption cross-section:
\be
\int d\nu_D \langle 1_{k,q} 0_D| \cT^\dagger | 0_\phi \nu_D \rangle
\langle 0_\phi \nu_D  | \cT | 1_{k,q}0_D \rangle = VT 2iq
\sigma_{\mathrm{abs}}~.
\ee
Here $\nu_D$ denotes an arbitrary state of the brane, $0_\phi$ ($0_D$)
the vacuum in the bulk (brane), and $2iq$ is the incident flux.

Altogether, unitarity implies
\be \label{sigma}
-2 q^{2l} \Ima \Delta_2^{(l)} \sum_{\vm }
C^{9\ldots 9}_{l\vm} C^{9\ldots 9}_{l\vm} =
2q \sigma_{\mathrm{abs}} + \frac{q^{4l+4}}{2(2\pi)^5} \cNl
 \sum_{\vm }
C^{9\ldots 9}_{l\vm} C^{9\ldots 9}_{l\vm} | \Delta_2^{(l)}|^2.
\ee
The absorption cross-section is related to the absorption probability
through\cite{gubser,gh}
\be
  \sigma_{\mathrm{abs}} = \frac{8\pi^2}{3q^5}(l+1)(l+2)^2(l+3)
 P_{\mathrm{abs}}~.
\ee
We can thus recast (\ref{sigma}) as an expression relating the absorption 
probability to the propagator,
\be \label{Pabs}
P_{\mathrm{abs}} = -2\Ima\,
\left(\frac{q^{2l+4}\Delta^{(l)}_2}{2^{l+5}\pi^2(l+1)(l+2)}\right)
-\left|\frac{q^{2l+4}\Delta^{(l)}_2}{2^{l+5}\pi^2(l+1)(l+2)}\right|^2~,
\ee
where we have used the values of $\cNl$ and $\CC$ given
in Appendix B. 
Inserting our explicit expression for $\Delta^{(l)}_{2}$, 
Eq.~(\ref{delta2lexp}), we obtain
\bea \label{Pabseval}
 P_{\mathrm{abs}} &=& - 2 \Ima\,
i\left(\frac{\chi-{1\over\chi}}{\eta\chi-{1\over\eta\chi}}-1\right)
                   -
\left|\frac{\chi-{1\over\chi}}{\eta\chi-{1\over\eta\chi}}
  -1\right|^2\\ \nonumber
&=& 1 - \left|\frac{\chi-{1\over\chi}}{\eta\chi-{1\over\eta\chi}}
\right|^2~.
\eea
As can be seen from (\ref{pabs}) and (\ref{AB}),
this is precisely the formula for the absorption as defined in the 
bulk. We have thus verified
that our two-point function
satisfies the optical theorem.

Note that elastic scattering and absorption processes
contribute to the total cross-section
at the same order in $g_s N$. At low energies elastic scattering
is suppressed, but
away from the conformal limit ($\omega R\to 0$) 
it has to be taken into account if one attempts
to reconstruct the propagator directly from the absorption probability.
This was overlooked in \cite{gh}.

 Another point to notice is that it is the full two-point function
 (\ref{delta2lexp})
 which contributes to (\ref{Pabseval}), including
 the contact (analytic) terms arising from the second term
 of (\ref{cotgformula}) or (\ref{cotgformula_l}). 
 We noted before that these terms are peculiar because
 they are imaginary for real values of $s$. 
 In field theory, analytic terms can arise
 from divergent diagrams when using cut-off regularization,
 but it is hard to see how they could have
 an imaginary part for real values of $s$.
 Since the problematic terms are analytic, 
 one is tempted to simply discard them. The puzzle, however, is that 
 the resulting two-point function would no longer satisfy the optical 
 theorem, Eq.~(\ref{Pabs}).
  
 We should stress that the appearance of this type of terms
is not unique to our approach:
the correlators derived in \cite{ms,mr} by means of
 an extrapolation of the GKPW recipe \cite{gkp,w} suffer
from the same difficulty. This was not noticed in those works, and 
neither was the tension between discarding these terms
and satisfying the 
optical theorem, since the authors
of \cite{ms,mr} did not attempt to establish
the relation between their proposed
two-point functions and the corresponding absorption probabilities.

The presence of these bizarre terms is related to a phenomenon
first pointed out by Stokes \cite{dingle}. The essential point is that
both the GKPW recipe and our prescription extract the 
subleading coefficient in the large-$r$ expansion of an expression 
of the type $E(s,\sqrt{s}r)$ (the expression in question is in our case
the curved-space propagator $G$; see (\ref{delta2})). 
The expansion is supposed to 
make sense for arbitrary complex values of $s$.
For negative 
real $s$, in particular,
the expansions of relevance to \cite{ms,mr} and the present paper
are all of the form
\be
E(s,\sqrt{s}r)\sim L(s){1\over r^{p}}e^{\sqrt{-s}r}
     +S(s){1\over r^{p}}e^{-\sqrt{-s}r}~,
\ee
with $p$ some constant. The prescriptions for two-point correlators
employed by the authors of \cite{ms,mr} and by us
essentially extract the subleading coefficient, $S(s)$.
To reproduce the correct holonomy
of the function $E$ upon encircling the origin
of the complex $s$-plane, the coefficients $L$ 
and $S$ develop peculiar properties. For instance, $S$ can appear
to be complex on the negative real $s$-axis, even if the 
function $E$ is manifestly real there\footnote{An example of this can 
be seen in the asymptotic expansion of the Bessel function $I_{\nu}(z)$ 
--- see, e.g., \cite{gr}.}.     

Taken at face value, the presence of imaginary analytic
terms in the propagator (\ref{delta2lexp}) would appear to indicate 
a breakdown of unitarity in the dual theory.
We believe, however, that 
there should exist a less radical interpretation. The simplest 
possibility is that these terms should be dropped. 
As we have  explained above, the
problem would then be
to understand why the resulting propagator fails to 
satisfy Eq.~(\ref{Pabs}).  Alternatively, one
could try to relax one of the 
assumptions employed in the derivation of Eq.~(\ref{delta2}).
For instance, one could take into account the back-reaction of the 
branes on the geometry, through the inclusion of tadpole diagrams.
This would entail the replacement of the
flat-space propagator $G_{0}$ in (\ref{delta2})
with some curved-space propagator $\tilde{G}$, which would
presumably differ from $G$ by the choice of boundary conditions.
The net effect of this or any other potential resolution should
be to multiply $\Delta_{2}$ by an overall complex `form factor'
$f(s)$, such that the resulting propagator has a standard analytic 
structure. At the same time, a compensating change should take place
in the kinematic factors appearing in Eq.~(\ref{Pabs}), 
to ensure that the 
optical theorem is still satisfied.

Since we have brought up the issue of back-reaction,
we wish to emphasize before closing this section that, 
in our opinion, the expectation that there could exist
a duality generalizing the 
AdS/CFT correspondence amounts to the hope that the sum over 
diagrams in the dual theory
with arbitrary numbers of loops will by itself reproduce the 
effects of the curved geometry. As explained in
\cite{verlinde,kv}, this hope ultimately rests on open string--
closed string duality (see \cite{pt} for a related discussion). 
At any rate, given the conceptual difficulties 
inherent in placing explicit D-branes in the curved
background they themselves generate, it is surely important 
to see how far one can develop
a duality involving the worldvolume theory
embedded in flat space.

\section{A One-Point Function and the UV/IR Correspondence}
\label{1ptsec}

A key ingredient of holography is the ability of the hologram to encode the
extra dimensions of the object it is meant to represent. 
In the standard AdS/CFT correspondence, the radial coordinate of AdS space
is mapped to a scale in the CFT--- phenomena at different scales in the
lower-dimensional theory correspond to phenomena at different radii in the
bulk. The larger the radius in AdS the smaller the length scale in the CFT.
This has come to be known as the UV/IR correspondence \cite{susswi, pp}, and
is a striking illustration of holography at work.

An interesting way to study the UV/IR correspondence is through an analysis
of the one-point function produced by an external source. As shown in \cite
{bklt} and further examined in \cite{dkk1}, a point source at fixed $%
r^{\prime }$ in the bulk of AdS space will manifest itself in the
holographic dual, through an associated expectation value, as a blob with a
radius that goes like $R^{2}/r^{\prime }$. The blob is extended if the
source is deep down in AdS, and concentrated if it is close to the boundary (%
$r^{\prime}\to\infty$).

The physical interpretation of the source is more transparent if instead of
a point source we consider a string originating at infinity and terminating
on the D3-branes. Each point on the string acts as a source for the field
and contributes to the one-point function. The total one-point function is
obtained by integrating along the string, and different parts of the string
dominate at different length scales in the holographic dual. The result
represents, from the point of view of the dual theory, the one-point
function in the presence of an external quark \cite{dkk1,cg}. While all of
this is well understood in the AdS case, it is the subject of the present
section to investigate what happens if we use the full D3-brane metric. We
will find that there are some interesting new phenomena in this more general
setup.

\begin{figure}[htb]
\centerline{\epsfxsize=6cm\epsfbox{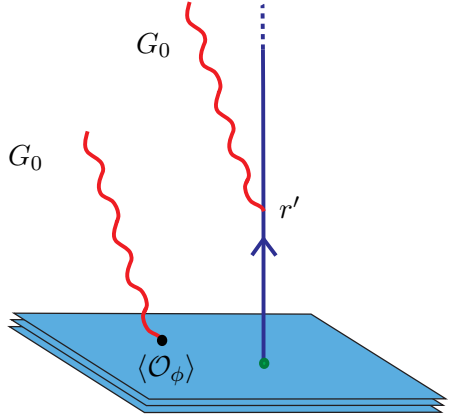}}
\begin{picture}(0,0)
  \put(47,39){\small $G_{0}$}
  \put(64,54){\small $G_{0}$}
  \put(83,32){\small $r'$}
  \put(64,11){\small $\left\langle {\mathcal{O}}_{\phi }\right\rangle$}
  \end{picture}
\vspace*{-0.7cm}
\caption{{\small The one-point function in the presence of an
external quark is computed with the aid of a string. Each point on the
string can be regarded as an individual source. A dilaton detected at
infinity may have originated from the string or from the D3-branes, through
the one-point function induced by the presence of the external source. See
text for discussion.}}
\end{figure}

We focus attention first on the case of a point source, located at some
radial position $r^{\prime}$. If desired, it can be regarded as an
infinitesimal segment of a string. The one-point function for the entire
string will be discussed a little bit later. For concreteness, we consider a
source for the dilaton s-wave. Proceeding as in the case of the two-point
function, we write down the one-point function with the help of 
\begin{equation}
G\left( r,r^{\prime }\right) =G_{0}\left( r,r^{\prime }\right) +G_{0}\left(
r,0\right) \Delta_{1,\,r^{\prime }}\left( q^{2}\right) ,  \label{g=g+g1d}
\end{equation}
where $G\left( r,r^{\prime }\right) $ (with $r\rightarrow \infty $) is the
supergravity propagation from the source out to infinity, divided into a
direct piece through flat space and a piece due to the induced one-point
function on the brane, $\Delta_{1,\,r^{\prime }}=\left\langle {\mathcal{O}}%
_{\phi }\right\rangle$, with ${\mathcal{O}}_{\phi}$ the operator given in (%
\ref{o}). The situation is summarized in Fig.~3 (which depicts the entire
string). The one-point function is thus computed using 
\begin{equation}
\Delta_{1,\,r^{\prime }}\left( q^{2}\right) =\lim_{r\rightarrow \infty }%
\frac{G\left( r,r^{\prime }\right) -G_{0}\left( r,r^{\prime }\right) }{%
G_{0}\left( r,0\right) }~.  \label{def1d}
\end{equation}
The subscript $r^{\prime }$ on the one-point function indicates the
dependence on the radial position of the source. Using the same notation and
definitions as in Section \ref{2ptsec}, the result can be written as 
\begin{equation}
\Delta_{1,\,r^{\prime }}\left( q^{2}\right) =\frac{4}{q^{2}r^{\prime 2}}%
\left( e^{i\left( \theta -\theta _{0}\right) }\frac{1}{A}H^{(1)}\left( \nu
,-\ln r^{\prime }/R\right) -\left( H_{2}^{(1)}\left( qr^{\prime }\right)
+H_{2}^{(2)}\left( qr^{\prime }\right) \right) \right) .  \label{1dmat}
\end{equation}

Let us now investigate this one-point function in the limit of large scales,
i.e., the IR\ regime in the holographic dual. To do this we consider
expression (\ref{1dmat}) in the limit of small momentum, $q\ll 1/R$, which
implies large distances $x$ after Fourier transforming. We begin by
considering $qr^{\prime }\ll 1$ but finite $qR^{2}/r^{\prime }$. This
includes the case with a source deep inside the AdS region (i.e. $r^{\prime
}\ll R$), and we therefore expect to recover the AdS result. We find that
the one-point function takes the form 
\begin{equation}
\Delta _{1,\,r^{\prime }}\left( q^{2}\right) =\frac{i\pi }{4q^{2}r^{\prime 2}%
}H_{2}^{(1)}\left( qR^{2}/r^{\prime }\right) -1,  \label{1dads}
\end{equation}
where we have used the fact that $A=-\frac{16i}{\pi R^{4}q^{4}}$ for small $%
q $. For the purpose of studying the UV/IR correspondence it is useful to
write down the one-point function for spacelike momenta $q^{2}=-p^{2}$,
where it becomes 
\begin{equation}
\Delta _{1,\,r^{\prime }}\left( -p^{2}\right) =\frac{p^{2}R^{4}}{2r^{\prime
2}}K_{2}\left( pR^{2}/r^{\prime }\right) -1.  \label{1deuc}
\end{equation}
The Fourier transform, 
\begin{equation}
\frac{1}{\left( 2\pi \right) ^{3}}\int d^{3}pe^{-i\vec{p}\cdot 
\vec{x}}\Delta _{1,\,r^{\prime }}\left( -p^{2}\right) =\frac{%
15R^{8}}{8\pi }\frac{r^{\prime -4}}{\left( x^{2}+R^{4}/r^{\prime 2}\right)
^{7/2}},  \label{fads}
\end{equation}
clearly shows how the radial position of a point source is reflected in the
holographic dual: the one-point function describes a blob whose width is of
order $R^{2}/r^{\prime}$. This is 
in accordance with the UV/IR correspondence.

The derivation above used $p\ll 1/R$ and $pr^{\prime }\ll 1$, and it is
clear that there will be small distance modifications when $x<R$. However,
 expression (\ref{fads}) 
 will be valid as long as $x\gg r^{\prime }$
even if $r^{\prime }$ is not small, i.e., even if it is
outside of the AdS region. On the other hand, if $r^{\prime }\sim x$ we
conclude that the one-point function may be corrected even at large
distances. To find out how, we need to consider the one-point function for $%
p\ll 1/R$, finite $pr^{\prime }$ and hence $pR^{2}/r^{\prime }\ll 1$. This
gives 
\begin{equation}
\Delta _{1,\,r^{\prime }}\left( -p^{2}\right) =\frac{16R^{4}p^{2}}{%
3r^{\prime 2}}K_{2}\left( pr^{\prime }\right) ,  \label{krprop}
\end{equation}
where we have made use of the fact that $\frac{B}{A}-1\sim \frac{2\pi
iR^{4}p^{4}}{3}$ for small $p$. After Fourier transforming in $p$ we get 
\begin{equation}
\frac{1}{\left( 2\pi \right) ^{3}}\int d^{3}pe^{-i\vec{p}\cdot 
\vec{x}}\Delta _{1,\,r^{\prime }}\left( -p^{2}\right) =\frac{%
20R^{4}}{\pi }\frac{1}{\left( x^{2}+r^{\prime 2}\right) ^{7/2}}  \label{fd3}
\end{equation}
for large $r^{\prime }$. Interestingly, the UV/IR correspondence is now
reversed, with the size of the blob increasing with the radial position. To
summarize: as a source is moved to ever larger values of $r^{\prime }$, the
blob first decreases in size, in accord with the usual UV/IR correspondence.
But as the source leaves the AdS region, the blob reaches a minimal size of
order $R$ and then starts to grow again. According to the standard UV/IR
correspondence, the region deep inside of AdS corresponds to the IR of the
holographic dual, while the region close to the boundary of AdS corresponds
to the UV. With AdS as part of a full three-brane background we can proceed
even further out, and according to the above reasoning we will again
encounter a region of space that will influence the IR behaviour of the
holographic dual. In the concluding section of the paper we will have more
to say on the nonstandard UV/IR properties of D3-brane holography, and
possible connections with non-commutative geometry.

In view of the modified UV/IR correspondence that we have observed in the
D3-brane background, it is also important to consider the result of
integrating over the source position, $r^{\prime }$, along the entire
string. As discussed above, this corresponds to determining the expectation
value of the operator dual to the dilaton s-wave, Eq.~(\ref{o}), in the
presence of an external quark. In the AdS limit we find that 
\begin{eqnarray}
\left\langle \frac{1}{4}F^{2}\right\rangle  &=&\frac{1}{\sqrt{2}\kappa}%
\left\langle {\mathcal{O}}_{\phi }\right\rangle =\frac{1}{4\pi \alpha
^{\prime }}\int dt^{\prime}dr^{\prime }\Delta _{1,\,r^{\prime }}\left( x\right) 
\label{1dstr} \\
&=&\frac{1}{4\pi \alpha ^{\prime }}\int dt^{\prime}dr^{\prime}
\int \frac{d^{4}k}{%
\left( 2\pi \right) ^{4}}\,e^{-ik_{0}t'-i\vec{k}\cdot\vec{x}}
\Delta _{1,\,r^{\prime }}\left(k_{0}^{2}-\vec{k}^{2}\right)   \nonumber \\
&=&\frac{1}{4\pi \alpha ^{\prime }}\int_{0}^{\infty }dr^{\prime }\int \frac{%
d^{3}k}{\left( 2\pi \right) ^{3}}e^{-i\vec{k}\cdot\vec{x}}
\Delta _{1,\,r^{\prime }}\left(-\vec{k}^{2}\right) 
=\frac{R^{2}}{16\pi ^{2}\alpha ^{\prime }x^{4}}=\frac{\sqrt{%
2g_{YM}^{2}N}}{16\pi ^{2}x^{4}}~,  \nonumber
\end{eqnarray}
where we have used that the time integral enforces $k_{0}=0$. Note the $%
\frac{1}{\sqrt{2}\kappa}$ in the first equality which cancels a similar
factor in the charge density of the string. These factors come about since
we are using a canonically normalized dilaton. The result of the calculation
is indeed in agreement with \cite{dkk1}, apart from the factor of two
mentioned in Section \ref{abssec}. 

What
happens if we take into account the portion of the string that lies outside
of the AdS region? Through the reversed UV/IR correspondence, this threatens
to change the large scale behaviour of the one-point function. However, for
a given scale $x\gg R$, one can easily estimate the modifications of $%
\left\langle {\mathcal{O}}_{\phi }\right\rangle $ due 
to portions of the string
with $r^{\prime}>x$,
by comparing the integral of (\ref{fads}) and (\ref{fd3}) 
from $r^{\prime}=x$ to $r^{\prime }=\infty $. This shows that any
modification will be at most of order $1/x^{6}$, and the coefficient of the
leading $1/x^{4}$ term will therefore not be modified.

It would also be interesting to consider the high energy or small distance
behaviour of the one-point function. For this we would need to investigate
the one-point function (\ref{1dmat}) in the limit where $q^{2}=-p^{2}$
becomes large and negative. This we have not done; in Section \ref{abssec}
we have already discussed some difficulties in extracting this limit.

\section{Higher Correlation Functions} \label{nptsec}

The prescription for two-point functions derived in
Section \ref{2ptsec} extracts 
information from the brane geometry by probing it with a bulk 
correlation
function evaluated at points at asymptotic distances from the brane. 
No relation was assumed between the radial positions of these points. For 
\( n \)-point functions we may formulate our prescription in the same 
way, but it is convenient and perhaps more natural to compute 
correlators between points on a common cutoff surface. Then we can 
formulate rules for 
computing brane \( n \)-point functions motivated by the same 
reasoning as before (see Fig.~4):
\newcounter{counter}   
\begin{list}{\arabic{counter})}{\usecounter{counter}}
\item Introduce a cutoff surface at  $r=\rl$. 
\item Draw all curved space Feynman diagrams 
for the $n$-point correlation function 
of the appropriate supergravity fields. Work in position space for 
$r$ and in momentum space for $x^{\mu}$. The $n$ external legs
have one end at $r=\rl$, 
and carry four-dimensional momenta $k_{1},\ldots,
k_{n}$. 
\item Replace the propagators $G(k_{i};\rl,r)$ associated with
\emph{external} legs by $G_{\Delta }(k_{i};\rl,r)$,
where $G_{\Delta }=G-G_{0}$, with $G_{0}$ the 
corresponding flat space propagator. 
\item Amputate each external leg, dividing
by $G_{0}(k_{i};\rl,0)$. 
\item Take the limit $\rl\to \infty$. 
\end{list}

Combining rules 3) and 4),
the factor on each external leg becomes
\begin{eqnarray}
\Delta_{1,\,r}(k^2) &=& \lim_{\rl\rightarrow \infty } 
	{ G_{\Delta}(k;\rl,r) \over G_0(k;\rl,0)} \\
	&=& {e^{i(\theta-\theta_0)}\phi_1(r)- 2 A \phi_{01}(r) 
	\over 2 A \phi_{01}(0)} \nonumber \\
	&=& {e^{i(\theta-\theta_0)}\phi_3(r)- \phi_{03}(r) 
	+ \frac{B}{A} e^{i(\theta-\theta_0)}\phi_2(r)- \phi_{02}(r) 
	\over   \phi_{01}(0)}~. \nonumber \label{Delta1}
\end{eqnarray}
The notation here is the same as in Section \ref{2ptsec}.
This factor effectively removes processes where the external leg is 
unaffected by the presence of the brane. We thus restrict attention 
to processes where all external lines touch
the brane, and from 
these we extract the physics on the brane.

\begin{figure}[htb]
\centerline{\epsfxsize=6cm\epsfbox{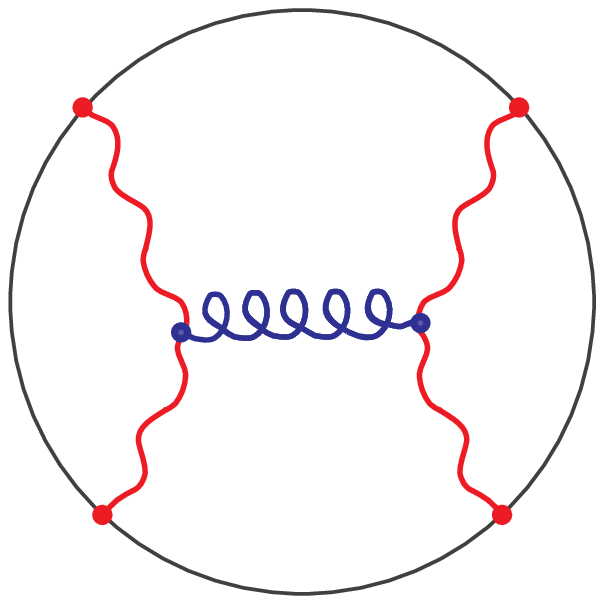}}
 \begin{picture}(0,0)
  \put(35,34){\small $r=\rl$}
  \put(75,26){$G$}
  \put(53,22){$G_{\Delta}$}
  \put(53,40){$G_{\Delta}$}
  \put(94,22){$G_{\Delta}$}
  \put(94,40){$G_{\Delta}$}
  \put(61,48){\vector(1,-2){3}}
  \put(64,45){$k_{1}$}
  \put(63,18){\vector(1,2){3}}
  \put(65,18){$k_{2}$}
  \put(92,48){\vector(-1,-2){3}}
  \put(85,45){$k_{3}$}
  \put(90,18){\vector(-1,2){3}}
  \put(84,18){$k_{4}$}  
 \end{picture}
 \vspace*{-0.5cm}
\caption{\small The prescription for $n$-point correlators: 
Introduce a cutoff surface at asymptotic radial coordinate $r=\rl$. 
In each of the
Feynman diagrams for the $n$-point correlation function 
of the appropriate supergravity fields, 
replace all curved space propagators $G$ associated to 
\emph{external} legs 
by $G_{\Delta }=G-G_{0}$. Amputate each 
external leg, dividing by $G_{0}$. See text for a more precise 
description.}
\end{figure}

As an example we write down the case of the three-point 
function
\begin{equation}
    \Delta_{3}(k_{1},k_{2},k_{3}) =
    \lim_{\rl \to \infty} \frac{
    \int_{}^{} \sqrt{-g}dr G_{\Delta }(k_{1};\rl,r) G_{\Delta }(k_{2};\rl,r) 
G_{\Delta }(k_{3};\rl,r) V_{3}} 
{G_{0}(k_{1};\rl,0) G_{0}(k_{2};\rl,0) G_{0}(k_{3};\rl,0)}~,
    \label{eq:3-point}
\end{equation}
obtained from the corresponding supergravity three-point function
\begin{equation}
    G_{3}(k_{1},k_{2},k_{3};r_{1},r_{2},r_{3}) =
    \int_{}^{} \sqrt{-g}dr G(k_{1};r_{1},r) G(k_{2};r_{2},r) 
G(k_{3};r_{3},r) V_{3}~,
    \label{eq:SUGRA3-point}
\end{equation}
where \( V_{3} \) is the supergravity three-vertex, 
which generally may involve the 
metric, momenta in the brane directions and derivatives in the 
transverse dimensions.

We now want to check that these \( n \)-point functions agree with known 
results in the conformal limit, which amounts to showing that the 
integrals reduce to AdS integrals in the low energy limit \(qR \to 0\). 
To do this more information on the solutions \(\phi_{i}\) and 
\(\phi_{i0} \) is needed. We give the arguments for the three-point 
function, but one can repeat the same steps for any tree 
diagram, at least for any diagram with interaction vertices directly 
connected to external legs. For six-point functions and higher there 
are other diagrams, and the argument is not complete in its present 
form.  
 
We may use the asymptotics of the Mathieu functions obtainable 
through the series expansion 
(\ref{eq:Asymp}) described in Appendix A, in the limit of  small \( q r \).
To facilitate comparison with the AdS literature we use Euclidean \( 
q^2=-p^2 \) and find
\begin{equation}
\Delta_{1,\,r}( - p^2 ) \rightarrow\frac{p^{2}R^{4}}{2r^{\prime 2}}K_{2}\left(
pR^{2}/r^{\prime }\right)  -1  \rightarrow - \left({p R^2 \over 2 r}\right)^2 . 
\label{EuclidLim}
\end{equation}
We note that \( \Delta_{1,\,r}( k^2 ) \) vanishes for \( r \gg pR^2 
\) (if \(p r \ll 1 \) ). As demonstrated below the contribution from 
the outer region, 
\(p r \geq 1 \), is suppressed for low energies.
Thus the effective upper limit of integration for the three-point function in 
Eq. (\ref{eq:3-point}) goes to zero in the low energy limit, and we 
may use the near horizon limit of the integration measure to find
\begin{eqnarray}
\lefteqn{\Delta_3^{CFT}(k_1,k_2,k_3) \sim \kappa R^2 \int dr r^3 V_{3 }
\Delta_{1,\,r}( q_1^2 )
	\Delta_{1,\,r}( q_2^2 ) \Delta_{1,\,r}( q_3^2 ) } \\
	&=&
	\kappa \frac{R^8}{8} \int dz z^{-5}\left(p_1^2 z^2 K_2(p_1 z) -2 \right)
	\left(p_2^2 z^2 K_2(p_2 z) -2 \right)
	\left(p_3^2 z^2 K_2(p_3 z) -2 \right)~, \nonumber
	\label{AdS3Point}
\end{eqnarray}
which scales as \( q^4 \) in the low energy limit. We recognize the AdS 
form of the three-point function with the bulk-to-boundary 
propagator \( p^2 z^2 K_2(p z) \). 
The subtractions found automatically in our approach may look 
unfamiliar, because they are not always explicitly mentioned. They 
are however necessary to obtain the final finite conformally 
invariant results given for instance in 
\cite{fmmr,magoo}. They are all 
independent of at least one momentum (or at least one relative distance, in 
position space). Just as in the case of the two-point function,
our approach 
automatically produces a finite answer, in contrast to the standard AdS 
calculations that require renormalization. Since the region of 
integration that contributes is entirely within the AdS region, any 
supergravity vertex \( V_{3} \) that yields a conformally invariant 
three-point function in the AdS case, will do so in our low energy limit, 
no matter how complicated it is. The three dilaton vertex,the 
dilaton-\({\rm Tr}F^2 \) coupling, our normalization of \(F\) and \( \kappa 
/R^4 \sim 1/N \) then give 
\begin{equation}
	\left\langle {\rm Tr}F^2 {\rm Tr}F^2 {\rm Tr}F^2 \right\rangle
	\sim \frac{1}{N}~,
\end{equation}
as expected.

To see why the region \( r \leq pR^2 \) dominates low energy behaviour 
we need a qualitative understanding of the wave solutions \( \phi_i(r) 
\). This may be obtained directly from their equation of motion 
\cite{gh}, written in terms of \( \rho = \ln \frac{r}{R} \),
\begin{equation}
	\left\{\partial_\rho^2  + 
	2(qR)^2 \cosh{2\rho} -4\right\} e^{2\rho} \phi_i(Re^{\rho})=0~,
\end{equation}
which can be thought of as a one dimensional Schroedinger 
equation with a potential barrier. For small \( qR \) there is 
oscillatory behaviour for \( 2r < qR \) and 
for \( qr >2 \). The waves tunnel through the barrier between the turning 
points with an 
approximate amplitude \( \phi(r) = C + DR^4/r^4 \) inside the barrier. 
For the solution \( \phi_1 \), which is purely ingoing for small \( r \), 
almost all the 
incident wave is reflected and the amplitude does not grow from the 
exterior region towards the interior region. This means that \( C \gg D 
\). For the exterior region we use Eqs. (\ref{eq:Asymp}), 
(\ref{Delta1}) and (\ref{bo}), which give
\begin{equation}
	I_{ext} \sim \int_{1/q} r^5 dr 
	q_1^4\Delta_2{(q_1^2)}
	q_2^4\Delta_2{(q_2^2)}
	q_3^4\Delta_2{(q_3^2)}
	\frac{H^{(1)}_2(q_1 r)}{(q_1 r)^2}\frac{H^{(1)}_2(q_2 r)}{(q_2 r)^2}
	\frac{H^{(1)}_2(q_3 r)}{(q_3 r)^2} \sim q^6~,
\end{equation}
using \( \Delta_2(q^2) = {\mathcal{O}}(q^0)\). The contribution from 
the barrier region 
is obtained by matching to the exterior solution at 
the exterior turning point: 
\begin{equation}
	I_{bar} \sim \int_R^{1/q} r^5 dr {q_1}^4 {q_2}^4 {q_3}^4 \sim q^6~,
\end{equation}
using also Eq.~(\ref{eq:nuAppr}).
We find as promised that the conformal 
result (\ref{AdS3Point}) from the interior region dominates at low 
energies.

\section{The Quark-Antiquark Potential} \label{wilsonsec}

In the standard AdS$_{5}$/CFT$_{4}$ correspondence, an external quark 
in SYM is dual to a string in the bulk of AdS space. This 
identification leads to a natural recipe \cite{reyee,juanwilson}
for computing Wilson loops in the strongly-coupled
gauge theory,\footnote{See also \cite{dgo} for a 
detailed discussion and refinement of the recipe.} 
which has been exploited 
to obtain numerous interesting results 
(see \cite{sonnenschein,magoo} for a review of some of them).
In the present 
section we wish to generalize the Wilson loop prescription of 
\cite{reyee,juanwilson} to our setting, and use it to
determine the 
quark-antiquark potential in the holographic theory. 

The motivation for relating strings in the bulk theory
to external sources in the four-dimensional theory is of 
course the same as in the AdS case.
Start with $N+1$ coincident D3-branes, and pull one brane out to 
a finite separation $\rl$, 
by giving a vacuum expectation value to the 
appropriate worldvolume scalar field. This breaks $U(N+1)\to 
U(N)\times U(1)$. A string connecting the solitary 
brane to the stack of $N$ branes represents a W-boson of the 
spontaneously broken gauge theory \cite{wittenbound}. In the
supergravity picture, the $N$ 
D3-branes are replaced by the black three-brane solution, so the
W-boson corresponds to a string extending from the solitary D3-brane
at $r=\rl$ down to the horizon at $r=0$.
In Section \ref{1ptsec} we have already made use of this
representation to determine the 
field around a point source in the dual theory.

We describe the dynamics of a fundamental string through the 
Nambu-Goto action
\be \label{ng}
S_{F}= -T_{F} \int d^{2}\sigma \sqrt{-g^{(i)}}~,
\ee
where $g^{(i)}$
is the pullback of the spacetime metric (\ref{d3metric})
to the string worldsheet, and
$T_{F}={1/ 2\pi\ls^{2}}$ is of course the string
tension. 
We make a static gauge choice $\sigma^{1}=r, \sigma^{2}=t$, 
and restrict attention to static
configurations of the form $X(r)$ (where $x$ is one of the spatial 
directions parallel to the D3-branes),
with the string pointing along a fixed $\bS^{5}$ 
direction. In most of the discussion
it will be convenient to use an 
inverted radial coordinate $z=R^{2}/r$.
The generic static solution,
\be \label{geodesic}
X(z;z_{m})=\pm\int_{z}^{z_{m}}dz\,
     \frac{z^{4}+R^{4}}{z^{2}\sqrt{z_{m}^{4}-z^{4}}},
\ee
describes a
string lying along a geodesic which 
starts and 
ends at the location of the probe D3-brane, $r=\rl$
($z=\zl$), 
and extends down to a minimum at $r=r_{m}$
(a maximum at $z=z_{m}$), 
as shown in Fig.~5a. 
We will eventually take $\rl\to\infty$,
to remove the probe brane.  The endpoints
of the string on this brane are separated by a distance 
\be \label{strsep}
\Delta X(\zl;z_{m})=2\int_{\zl}^{z_{m}}dz\,
     \frac{z^{4}+R^{4}}{z^{2}\sqrt{z_{m}^{4}-z^{4}}}.
\ee
For large $r$ the ambient
space becomes flat, so the string of course just lies along a straight
line, with slope  $\Delta x/\Delta r = R^{2}/z_{m}^{2}$.
The total energy of the string is
\be \label{strenergy}
U(\zl;z_{m})=2 T_{F}R^{2}\sqrt{R^{4}+z_{m}^{4}}
   \int_{\zl}^{z_{m}}
    \frac{dz}{z^{2}\sqrt{z_{m}^{4}-z^{4}}}~.
\ee

\begin{figure}[htb]
\centerline{\epsfxsize=\textwidth\epsfbox{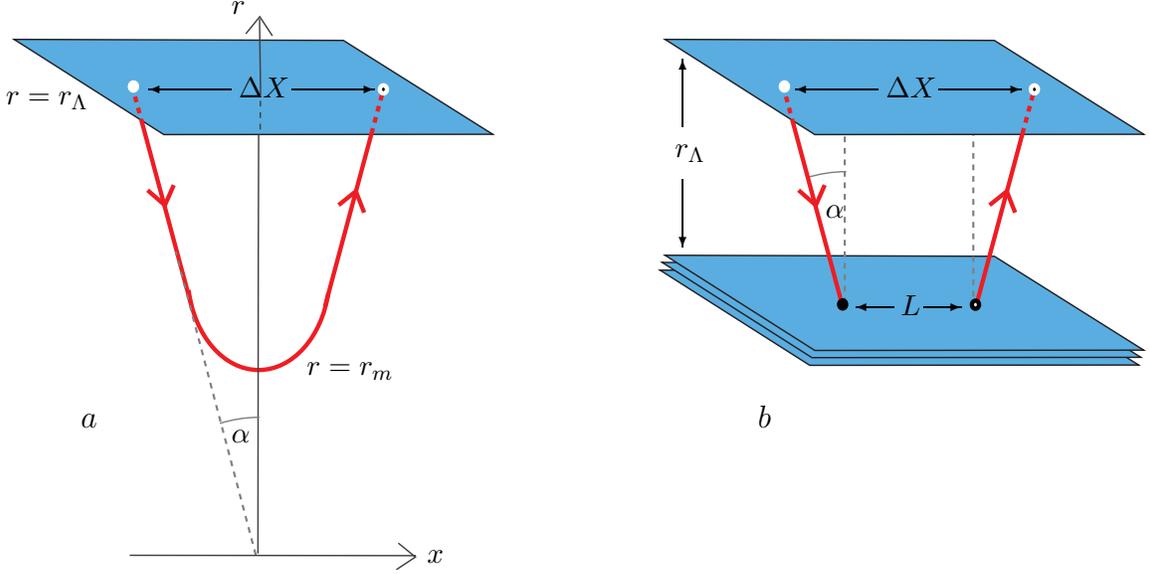}}
 \begin{picture}(0,0)
  \put(10,25){$a$}
  \put(100,25){$b$}
  \put(30,80){\small $r$}
  \put(56,7){\small $x$}
  \put(30,23){\small $\alpha$}
  \put(40,32){\small $r=r_{m}$}
  \put(31,69){\small $\Delta X$}
  \put(30,70){\vector(-1,0){11}}
  \put(38,70){\vector(1,0){11}}
  \put(0,68){\small $r=\rl$}
  \put(109,53){\small $\alpha$}
  \put(117,69){\small $\Delta X$}
  \put(116,70){\vector(-1,0){11}}
  \put(124,70){\vector(1,0){11}}
  \put(89,61){\small $\rl$}
  \put(90,65){\vector(0,1){9}}
  \put(90,58){\vector(0,-1){9}}
  \put(119,40){\small $L$}
  \put(118,41){\vector(-1,0){5}}
  \put(122,41){\vector(1,0){5}}
 \end{picture}
 \vspace*{-0.5cm}
\caption{\small a) `Hanging string' lying along a geodesic of 
the curved three-brane geometry. The endpoints of the string
lie on a probe D3-brane at $r=\rl$. This is the supergravity
realization of a  W-$\bar{\mbox{W}}$ pair in the 
$U(N)\times U(1)$ gauge theory.
b) The same system in the `branes + flat space' picture: 
two strings of opposite orientation connect the stack of $N$ 
D3-branes to a solitary brane a distance $\rl$ away. 
The endpoints of the strings on the stack of D3-branes
constitute a quark-antiquark pair. See text for 
discussion.}
\end{figure}

The string we have just described
corresponds to a W-$\bar{\mbox{W}}$ pair in the 
$U(N)\times U(1)$ theory, i.e., a quark-antiquark pair
from the perspective of the $U(N)$ theory. As a simple
check, notice
that if we send $z_{m}\to\infty$ holding $\zl$ 
fixed, (\ref{strenergy}) reduces to $2T_{F}R^{2}/\zl=2T_{F}\rl$,
which is the correct energy for two infinitely separated 
W-bosons\footnote{Incidentally, notice that this equality between
the total energy of a purely radial string in the curved background
and the corresponding W-boson provides a canonical way to identify 
the radial coordinates in the curved and flat backgrounds. This is 
significant for the prescription for correlation functions presented 
in Sections \ref{2ptsec} and \ref{nptsec}, which involves a comparison of 
the curved and flat space propagators.}.
In the `D-branes $+$ flat 
space' picture, the situation is as portrayed in Fig.~5b. The stack 
of $N$ D3-branes and the solitary brane are separated by a distance 
$\rl$, with two strings of opposite orientation running between them. 
The endpoints of these strings which lie on the $N$ D3-branes 
constitute a quark-antiquark pair in the worldvolume theory, and so
attract one another. As a result of this attraction, the strings are 
tilted by an angle $\alpha=\arctan (R^{2}/z_{m}^{2})$. The endpoints 
on the probe brane are held in place by an external agent
which enforces
the appropriate Dirichlet boundary conditions \cite{dgo}. 
Because of the tilt, the separation
$L$ between the quark and the
antiquark is (for large $\rl$) much smaller
than that between the endpoints on the solitary brane. As seen in
Fig.~5b, the two distances are 
related by
\be \label{sepdiff}
L=\Delta X(\zl;z_{m}) - 2 \rl\tan\alpha=
  \Delta X(\zl;z_{m}) - {2R^{4}\over \zl z_{m}^{2}}~.
\ee

We are now ready to compute the quark-antiquark potential. Since we 
are interested in taking the limit $\zl\to 0$ to remove the probe 
brane to infinity, we first carry out a Laurent expansion of 
(\ref{strsep}) about $\zl=0$, writing 
\bea 
\Delta X(\zl;z_{m})&=& 
     R^{4}\int_{\zl}^{z_{m}}
     \frac{2 dz}{z^{2}\sqrt{z_{m}^{4}-z^{4}}}
     +\int_{\zl}^{z_{m}}
     \frac{2 z^{2}dz}{\sqrt{z_{m}^{4}-z^{4}}}.  \\
{}&=&{R^{4}\over z_{m}^{3}}\left\{\int_{0}^{1}{d\zeta \over \zeta^{2}}
     \left[{2\over\sqrt{1-\zeta^{4}}}-2\right] +2{z_{m}\over \zl}-2
     \right\}
     +z_{m}\int_{0}^{1}\frac{2\zeta^{2}d\zeta}{\sqrt{1-\zeta^{4}}}
     +\ldots, \nonumber
\eea
where we omit terms involving positive powers of $\zl$. The two 
integrals can be carried out analytically, yielding $2-c$ and $c$, 
respectively, with 
$c=(2\pi)^{3/2}/[\Gamma(1/4)]^{2}\simeq 1.198$. We are thus left with
\be \label{strseplaurent}
\Delta X(\zl;z_{m})={2R^{4}\over 
\zl z_{m}^{3}}+c\left(z_{m}-{R^{4}\over z_{m}^{3}}\right)+\ldots
\ee
Using this in (\ref{sepdiff}) we obtain a relation
between the quark-antiquark separation $L$ 
and the geodesic parameter 
$z_{m}$,
\be \label{L}
L(z_{m})=c\left(z_{m}-{R^{4}\over z_{m}^{3}}\right),
\ee
which is perfectly well-defined in the limit $\zl\to 0$.

We next Laurent-expand (\ref{strenergy}),
\bea \label{strenergylaurent} 
U(\zl;z_{m})&=& {T_{F}R^{2}\over z_{m}^{3}}\sqrt{R^{4}+z_{m}^{4}}
    \left\{\int_{0}^{1}{d\zeta \over \zeta^{2}}
     \left[{2\over\sqrt{1-\zeta^{4}}}-2\right] +2{z_{m}\over \zl}-2
     \right\}+\ldots \nonumber\\
{}&=&2T_{F}{R^{2}\over \zl}{\sqrt{R^{4}+z_{m}^{4}}\over z_{m}^{2}}
     -c{T_{F}R^{2}\over z_{m}^{3}}\sqrt{R^{4}+z_{m}^{4}}+\ldots
\eea
The leading term diverges in the limit $\zl\to 0$, but it is 
clearly just the energy $2T_{F}\rl\sec\alpha$
of the two straight strings in Fig.~5b. We
are interested only in the energy $E$ which arises
from the  $U(N)$ interaction between
the quark and antiquark, so
we subtract this leading term and 
obtain\footnote{Just like in the AdS case,
the energy associated with the $U(1)$ interaction between the 
string endpoints on the solitary brane is negligible in the strong 
't~Hooft coupling regime.} 
\be \label{E}
E(z_{m})=-c{T_{F}R^{2}\over z_{m}^{3}}\sqrt{R^{4}+z_{m}^{4}}~.
\ee
Eqs.~(\ref{L}) and (\ref{E}) give the quark-antiquark potential $E(L)$ in 
implicit form. The potential is plotted
in Fig.~6. 

\vspace{1cm}

\begin{figure}[htb]
\centerline{\epsfxsize=7cm\epsfbox{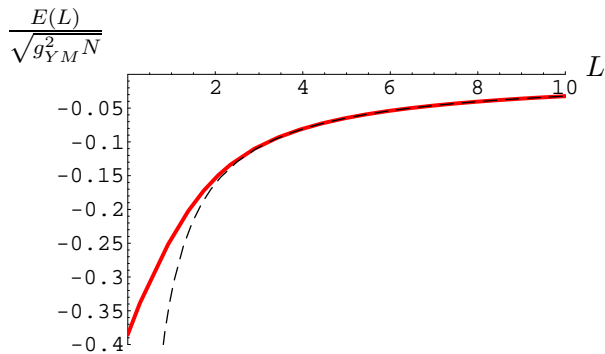}}
 \begin{picture}(0,0)
  \put(35,50){\small ${E(L)\over\sqrt{g_{YM}^{2}N}}$}
  \put(112,45){\small $L$}
 \end{picture}
 \vspace*{-0.7cm}
\caption{\small Quark-antiquark potential $E(L)$
in the holographic theory (in $R$=1 units). 
For large $L$, the potential agrees with that of
$\cN=4$ SYM (dashed line). For short
distances there is confining behaviour.}
\end{figure}

For large $q\bar{q}$ separation, $L\gg R$, both (\ref{L}) and (\ref{E}) 
reduce to the corresponding AdS relations \cite{reyee,juanwilson}, so 
as required the IR limit of the potential coincides with 
the conformal $\cN=4$ potential, 
\be \label{irpot}
E_{IR}(L)=-{c^{2}\over \sqrt{2}\pi}
          {\sqrt{g_{YM}^{2}N}\over L}~.
\ee
Here we have made use of the relation 
$R^{4}=2g_{YM}^{2}N\ls^{4}$ to 
write the result exclusively in terms of gauge theory quantities.

At the opposite extreme, notice from (\ref{L}) that, surprisingly,
$L\to 0$ as $z_{m}\to R$. Geodesics with $z_{m}<R$ thus appear to give 
no direct information about the quark-antiquark interaction in the 
$U(N)$ theory. Since these geodesics are completely outside the throat 
region $z\ge R$, this is yet
another indication that the flat space 
region is in a sense
left out of the holographic theory. This is 
as expected
from the decoupling argument of Section \ref{backgroundsec},
and is indeed
consistent with what we have found for correlation functions
in Sections \ref{2ptsec} through \ref{nptsec}.  
The short-distance
quark-antiquark potential is
\be \label{uvpot}
E_{UV}(L)=-{c\sqrt{g_{YM}^{2}N}\over \pi R}+ 
    {\sqrt{g_{YM}^{2}N}\over 2\pi R^{2}}L~.
\ee
This expression is again written only in terms of parameters of the
gauge theory,\footnote{Recall that $R$ itself is present
as a parameter in
the gauge theory away from the conformal limit, as seen in
(\ref{conjecture}).} 
and displays confining behaviour with a confining
string tension $\sigma=\sqrt{g_{YM}^{2}N}/ 2\pi R^{2}$. 
{}From the supergravity perspective 
this result is not surprising, as $\sigma=T_{F}/\sqrt{2}$
is simply the tension of a 
fundamental string located at $r=R$.

{}From (\ref{L}) and (\ref{E}) it follows that the quark and antiquark 
attract one another with a force
\be \label{forceN}
\frac{\p E}{\p L}=\frac{T_{F}R^{2}}{\sqrt{R^{4}+z_{m}^{4}}}~.
\ee
As a consistency check on our approach, we note that this equals 
the force $\p U/\p \Delta X$
obtained from the unsubtracted expressions 
(\ref{strsep}) and (\ref{strenergy}) in the $\zl\to 0$ limit.

\section{The Baryon} \label{baryonsec}

In the preceding section we have seen
how the AdS picture of a quark-antiquark pair can be 
generalized away from the conformal limit.
We now wish to point out that it is also
possible to extend the AdS description of the baryon
\cite{wbaryon,groguri,imamura,cgs} to our setting.  

In the AdS$_{5}$/SYM$_{4}$ correspondence, the 
$SU(N)$ baryon (the color-neutral coupling of $N$ external 
quarks) is dual to $N$ fundamental strings stretching from the 
AdS boundary to a D5-brane wrapped around $\bS^{5}$
\cite{wbaryon,groguri}. To provide
a thorough description of this system it is necessary to
analyse the full worldvolume action for the D5-brane 
embedded
in AdS$_{5}\times\bS^{5}$. The baryon is then seen to
be realized as a particular class of BPS D5-brane
embeddings \cite{imamura,cgs,gomis,gomis2}, 
in which the $N$ strings appear as 
Born-Infeld string tubes \cite{calmal,gibb}.

In our case, then, to obtain a picture of the baryon
we must consider a D5-brane embedded 
in the full three-brane background.  It was shown 
in \cite{cgs,gomis,camino} that
this system has a one-parameter family of BPS solutions of the form
$\Z=\Z(\X;\Zl)$, 
where $\Z=-r\cos\theta$, $\X=r\sin\theta$,
with $\theta$ the polar angle on $\bS^{5}$. 
These embeddings have a flat portion and 
(for $\Zl>0$) a tubular region,
which as explained in \cite{cgs} represent a flat
D5-brane located at $\Z=\Zl$, and
a bundle of $N$ Born-Infeld strings, respectively (see Fig.~7).

\begin{figure}[htb]
\centerline{\epsfxsize=7cm
\epsfbox{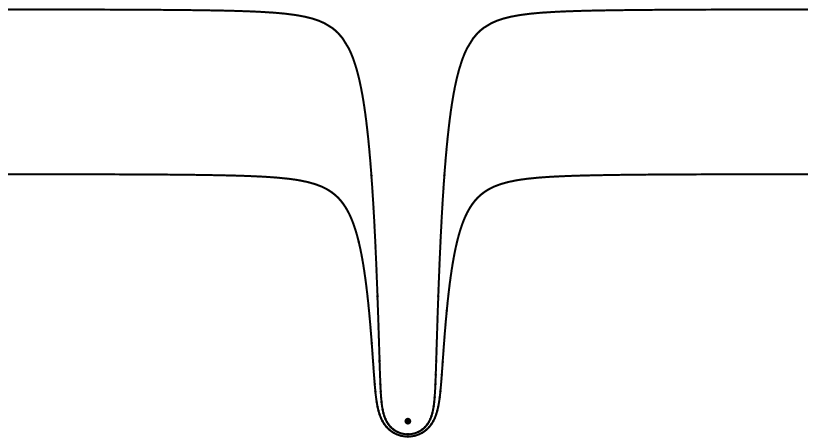}}
 \begin{picture}(0,0)
  \put(113,31){\small $\Zl=6R$}
  \put(113,45){\small $\Zl=10R$}
 \end{picture}
 \vspace*{-0.5cm}
\caption{\small Family of D5-branes embedded in the 
full D3-brane geometry. The solutions  extend
along $r$ and the $\bS^{5}$ directions,
and lie
at a fixed position along $\bf x$. 
They 
are parametrized as 
$\Z=\Z(\X;\Zl)$, 
where $\Z=-r\cos\theta$, $\X=r\sin\theta$, with
$\theta$ the polar angle on $\bS^{5}$. 
$\Zl$ denotes the vertical
position of the flat region of 
the D5-brane. 
The tubular portion of the
brane, in the limit $\Zl\to\infty$, is dual to the baryon
of the holographic theory.}
\end{figure}

We propose that these solutions are dual to
the baryon of the holographic theory,
with the understanding that 
the flat D5-brane at $\Z=\Zl$
plays the same role as the solitary
D3-brane at $r=\rl$ in the quark-antiquark case, and should
therefore be removed by taking $\Zl\to\infty$. For large 
$\Zl$, the lower portion of the tube is close to $r=0$,
and coincides with the AdS baryon embedding. This is a
necessary condition
for the large-distance field of the baryon
to reduce to that of the SYM 
theory\footnote{The latter was computed
in \cite{cg}. It would be interesting to repeat that calculation in 
our non-conformal setting, employing reasoning similar to that of
Section \ref{1ptsec}.}.
In the limit 
$\Zl\to\infty$, the Born-Infeld
tube becomes infinitely thin.

It was demonstrated
in \cite{cgs,gomis} 
that the energy of the tube (obtained
from the full energy by subtracting the infinite contribution
of the flat portion) is exactly equal to the energy of $N$ 
fundamental strings of length $\Zl$. Since these strings
represent the baryon's constituent quarks, it follows that
the D3 baryon (just like its AdS counterpart)
has zero binding energy, as expected from the BPS 
character of the configuration.

\section{Discussion} \label{lastsec}

Based on the assumption that there exists
\emph{some} theory which holographs the full
asymptotically flat three-brane background,
in this paper we have developed methods for computing
quantities in the dual theory in terms of
supergravity. 
Concretely, in Section \ref{nptsec} we have presented
a calculational recipe for arbitrary $n$-point 
correlators in the holographic dual. 
The particular case of 
two-point functions of the operators $\cO_{\phi}$
coupling to dilaton partial waves was analysed at length in
Sections \ref{2ptsec} and \ref{abssec}. 
Three-point functions were discussed
in Section \ref{nptsec}, and a
one-point function in the presence of a source was examined in 
Section \ref{1ptsec}. Additionally, 
in Section \ref{wilsonsec} we have employed a `hanging' string to 
determine the potential energy of an external quark-antiquark pair 
in the dual theory, and
in Section \ref{baryonsec} we have commented on the 
representation of baryons as deformed D5-branes.  

In addition to its derivation, we have presented several
non-trivial checks of our recipe for correlation functions. 
First, the correlators we obtain reduce to
   $\cN=4$ SYM correlators in the extreme low-energy limit. 
Second, the two-point functions of operators dual to dilaton
   partial waves are related to the corresponding exact 
   absorption probabilities \cite{gh}
   through the appropriate optical theorem.
Third, our prescription incorporates a natural
subtraction and amputation procedure which 
automatically removes potential UV divergences and renders
the correlators well-defined.

Despite these nice features, our results are not entirely 
satisfactory. The 
correlators we obtain appear to include 
terms which are analytic functions of the momenta with 
\emph{imaginary} coefficients.  
As explained in Section
\ref{abssec}, their presence is related to the 
so-called Stokes phenomenon \cite{dingle},
which in some cases
forces subleading coefficients 
of the asymptotic expansion of a real function to take on complex
values.
We should emphasize that the appearance of such terms
is not unique to our approach:
the two-point functions derived in \cite{ms,mr} using
the GKPW recipe \cite{gkp,w} suffer
from the same problem (a fact which was not noticed in those works).
Given that the problematic terms are analytic in momentum-space,
and therefore amount 
to contact terms, one is tempted to simply drop them 
from the correlators. This would indeed yield $n$-point functions with 
sensible analytic structure. The only problem is that these same 
contact terms appear to be necessary for two-point
functions to be correctly
related to the corresponding absorption probabilities, as
dictated by the optical theorem.
We have discussed
this puzzle in Section \ref{abssec},
as well as some of its possible resolutions.

Our work was motivated by a series of 
recent papers \cite{ghkk,dealwis,gh,intri}
addressing the possible generalization of the 
AdS$_{5}$/SYM$_{4}$ correspondence away from the
conformal limit $\omega R\to 0$. As explained in Section 2,
the authors of 
\cite{ghkk,dealwis,gh} espouse the view
that physics on the full three-brane 
background is encoded in the D3-brane worldvolume action (an insight 
which was implicit already in \cite{klebabs}).  Through 
considerations of symmetries and large anomalous dimensions,
Gubser and Hashimoto \cite{gh}
were led to conjecture that, for strong 't~Hooft 
coupling, this worldvolume 
theory is simply $\cN=4$ SYM deformed in the infrared
by a specific dimension-eight 
operator, as indicated in (\ref{conjecture}).
They advocate a duality between this theory and supergravity in the 
full three-brane background, which is supposed to hold
in the limit $\gs\to 0$, $\gs N\to\infty$.

We wish to stress that
our work does not rely on the specific form of the conjecture
of \cite{gh}. Since our calculations are based on the supergravity
side of the duality relation,
the physical quantities we compute pertain to whatever theory 
turns out to be dual to the curved space description, even if it
is not precisely of the form (\ref{conjecture}).
 We regard our work as a step towards the more precise specification 
 of this theory, and it is an outstanding challenge
 to reproduce our results through an explicit
 field theory calculation. 
 
On the other hand, it should be noted 
that our approach is in accord with the perspective of
\cite{ghkk,dealwis,gh} in two important respects:
first, throughout the paper we explicitly assume
that the worldvolume theory of D3-branes
(or at least some localized
$(3+1)$-dimensional object)
is relevant for the duality; second, for the most part we restrict 
attention to the limit $\gs\to 0$. 
These two assumptions play a role both in our  
derivation of the recipe for two-point functions in Section \ref{2ptsec}, 
and in our formulation of the optical theorem in Section \ref{abssec}. 

The duality we have studied in this
paper operates between
two alternative descriptions of physics in 
the presence of a system of $N\gg 1$ 
branes: on the one hand, supergravity on the curved three-brane 
background; on the other hand, the D3-brane worldvolume theory coupled to 
supergravity in the bulk of flat $(9+1)$-dimensional space. We are 
interested in this system  at finite $\omega R$, since $\omega R\to 0$ 
is just the usual Maldacena limit \cite{malda}.
An important aspect that follows from our analysis
(see Section \ref{2ptsec})
is that, for large $\gs N$, the worldvolume theory 
\emph{does not decouple} from the bulk even if $\gs\to 0$. As a 
result, $n$-point correlators of operators in the lower-dimensional theory
receive contributions from virtual particles propagating 
in the higher-dimensional space, which cannot be 
disentangled from those 
entirely confined to the brane. 
As explained in Section 
\ref{abssec}, such higher-dimensional
processes also have an impact on the form of
the optical theorem connecting two-point functions to absorption 
probabilities, a fact which was overlooked in \cite{gh}.

The lack of a complete
decoupling should not be mistaken
for the absence of a duality.
It does however indicate that, in contrast with the AdS/CFT case,
the `gauge theory' 
side of the duality in question
necessarily involves \emph{both} the 
worldvolume theory \emph{and} supergravity in flat space. 
Of course, as $\gs\to 0$ the latter theory 
becomes trivial--- free fields propagating in $(9+1)$-dimensional 
flat space. Roughly speaking, this trivial part of the system describes
the free fields on the asymptotically flat part of the three-brane 
geometry, while the worldvolume theory encodes all the non-trivial 
physics in the throat (which includes much more than the near-horizon 
geometry relevant to the standard AdS/CFT correspondence \cite{malda}).
If correct, this is undoubtedly a very profound statement.
In particular, it still seems
appropriate to us to speak of `holography', given that the non-trivial 
aspects of a higher-dimensional theory are encoded in the dynamics
of a theory which 
is essentially four-dimensional. 

Intriligator \cite{intri}
has argued that the duality conjectured by Gubser and Hashimoto 
should in fact be expected to hold for arbitrary $\gs$ and $N$. 
In Section \ref{backgroundsec}
we have reviewed his arguments
and commented on the problematic aspects
of his proposal.
Here we wish to emphasize that
Intriligator arrived at his conjecture
by means of scaling and non-renormalization arguments which do
not explicitly bring the two alternative 
D-brane descriptions into play.  While he asserts that
the dual theory is of the form (\ref{conjecture}), he
does not identify it
with the D3-brane worldvolume theory. 
This could perhaps be viewed 
as a weakness of his proposal, but 
it enables him to take the view that
the four-dimensional theory 
whose Lagrangian is
(\ref{conjecture}), is by itself (without any coupling to a 
higher-dimensional theory) 
dual to the \emph{entire}
three-brane background. 
While we approach the problem from a different perspective, 
the interesting question remains
whether there could exist a purely four-dimensional theory 
which holographs the full asymptotically flat geometry. From our
analysis it is clear that such a theory 
would necessarily be more than a  
pure D3-brane worldvolume theory, 
for it would have to summarize
the effective theory on the 
branes,
free supergravity in $(9+1)$-dimensional
flat space, and the interaction between them. 
In such a theory our correlators would 
be seen to
follow from a strictly four-dimensional calculation.

In this connection, we cannot resist commenting on the  
similarity between certain aspects of our results and some recent
analyses of non-commutative field theories.
A first point to notice is the potential connection of our work and
that of Maldacena and Russo \cite{mr},
who studied 
certain supergravity backgrounds which are the putative 
duals of large $N$ 
non-commutative gauge theories (this was first done in
\cite{hi}). 
The backgrounds in question are obtained as decoupling limits
of the solution describing $N$ D3-branes in the presence
of a constant $B$ field
--- the same limit which on the gauge theory 
side gives rise to the non-commutative description \cite{sw}.
Among other things, the authors of \cite{mr} computed 
a two-point 
correlator of a certain component of the gauge theory energy-momentum 
tensor,
employing an extrapolation of the GKPW recipe \cite{gkp,w}. 
Interestingly, the relevant supergravity field (restricted
to be independent of time and one spatial direction) satisfies an 
equation which is identical to that of a dilaton propagating on the 
full (zero $B$ field) 
asymptotically flat D3-brane background! As a result, the 
correlator obtained in \cite{mr}
is essentially the same as that computed by us in Section 
\ref{2ptsec}, and expressed in 
Eq.~(\ref{delta2exp}).  Maldacena and Russo also determined
the shape and energy of strings lying along geodesics of their 
geometry; the relevant equations happen to have the same form as 
those considered by us in our study of the quark-antiquark potential 
(see Section \ref{wilsonsec}). 
It is too soon to tell whether these 
remarkable similarities are merely accidental or
indicate some underlying connection.
It should be emphasized that the backgrounds studied in \cite{hi,mr}
are completely different
from ours: not only is there a non-vanishing $B$ field (whose magnitude is
in fact taken to diverge in the decoupling limit), 
but also a dilaton with non-trivial dependence on the radial 
coordinate. Still, it is interesting to note that, 
as pointed out in \cite{hi},
the Einstein frame metric in one case is asymptotically flat.

There is yet another feature of our results which 
is suggestive of a relation to the non-commutative case: the presence,
in what is otherwise a four-dimensional theory, of virtual 
particles propagating in a higher-dimensional space. This 
closely resembles a phenomenon encountered in recent studies of 
non-commutative perturbative dynamics \cite{mvrs,vrs} (see also 
\cite{abk}--\cite{gll}). In that context,
the basic 
observation is that while the non-commutativity removes
the usual UV field-theoretic divergences, it gives rise 
(even in a massive theory) to 
peculiar IR divergences.  
The latter represent novel
long-distance effects which 
in fact arise from 
the short-distance degrees of freedom---
an intriguing phenomenon known as 
UV/IR mixing \cite{mvrs,vrs}.  To capture the correct long-distance
physics in a Wilsonian description of the theory 
with an explicit UV cutoff, it is then necessary to introduce 
additional fields which
reproduce the unfamiliar IR singularities. Some of 
these fields have propagators which can be naturally 
interpreted as describing particle
propagation in a higher-dimensional space.
This suggests an analogy with string theory,  associating the 
lower-dimensional fields with open strings attached to a brane, and 
the higher-dimensional fields with closed strings which can propagate 
outside the brane\footnote{In situations where the non-commutative 
field theory is a limit of string theory \cite{sw}, 
this could perhaps be 
more than just an analogy \cite{vrs,kl} 
(see however \cite{ad,gkmrs,lm}).}.
There is thus a clear parallel with the `branes $+$ flat space'
description of our system, which incorporates
worldvolume and bulk degrees of freedom  
in manifest correspondence with massless modes of 
open and closed strings, respectively.

The imprint of UV/IR mixing is
seen also in the context of
our discussion of the UV/IR correspondence 
in Section \ref{1ptsec}. As explained there, a point-like source 
located at radial position $r=r'$ in the bulk makes its
primary contribution to 
a one-point function in the dual theory
at a length scale $L(r')$ which is essentially the maximum 
of $R^{2}/r'$ and $r'$. This means in particular that the 
$r'\to\infty$ region, which would be naturally
interpreted as a high-energy region in 
the holographic dual, can in fact give rise 
to the large-distance effects. 

We reiterate that all similarities with the non-commutative case
might amount to no more than an intriguing 
analogy. It is however tempting to speculate that 
a)~perhaps there exists a \emph{purely}
four-dimensional
theory which holographs the entire three-brane 
background (flat region included), and 
b)~possibly this theory incorporates some type of non-commutativity. 
The dual description that we have 
directly scrutinized --- the one which follows directly from D-brane 
reasoning --- could then be regarded as a reformulation of the 
non-commutative dual in terms of two coupled components:
a Wilsonian worldvolume effective 
action (which according to \cite{gh} is
given by (\ref{conjecture}), with a cutoff of order $1/R$), 
and the 
$(9+1)$-dimensional flat space supergravity action, encoding the
additional long-distance effects. This interpretation
would be in line with the fact that,
as explained in Sections \ref{abssec} and \ref{1ptsec}, 
there is a subtlety in
extracting low-energy
results from the supergravity description. If we 
regard AdS space as the $r\to 0$ portion of the full three-brane 
geometry, carry out calculations in the complete
background, and then take the low-energy limit, the results
do not necessarily agree with those obtained in a space 
which is purely AdS from the beginning. 
In the `branes $+$ flat space' 
side of the duality, the issue is whether or not
one includes the contribution of the virtual particles
which propagate off the branes. 
In the hypothetical strictly four-dimensional description, this would 
correspond to the statement that the non-commutative theory does 
not reduce to the naive commutative description at low energies.

Before closing, we wish to remark that
our prescription for correlation functions is based on such simple 
considerations that it seems possible to generalize it to arbitrary 
asymptotically flat backgrounds. Of course, in the general case one 
is no longer guided by D-brane intuition, so the attempt to describe
the curved background in terms of a lower-dimensional effective theory
embedded in a higher-dimensional flat space is more of a guess---
even if it is still in consonance
with the spirit of \cite{thooft,susskind}.
But it is the very fact that it would take us
into uncharted 
terrain which could possibly
make this investigation worth our while.

\section{Acknowledgements}

We are grateful to Nadav Drukker,
Ansar Fayyazuddin, Hans Hansson, Esko Keski-Vakkuri,
Subir Mukho\-padhyay, {\O}yvind Tafjord,
and  Staffan Yngve for useful discussions.
We would also like to thank
Karen Ram\'\i rez for help with the figures.
The work of U.D., A.G., 
and B.S. was supported by the Swedish Natural 
Science Research Council (NFR), and that of M.K. by
the Swedish Foundation for International Cooperation 
in Research and Higher Education (STINT).

\section*{Appendix A}

In this appendix we review some properties of Mathieu functions following
\cite{trascfcn,gh,mmlz}. 
In \cite{gh} it was realized that the radial equation
for a massless field moving in the D3-brane background,
\be
\left[ \frac{d^2}{dr^2}+\frac{5}{r}\frac{d}{dr}-\frac{l(l+4)}{r^2} +
q^2 \left(1+\frac{R^4}{r^4}\right)\right]\phi(r) = 0,
\ee
can be recast as  Mathieu's equation
\be \label{mathieueqn}
\left[\frac{d^2}{d\rv^2}+2(qR)^2 \cosh(2\rv) - (l+2)^2\right] 
\psi(\rv)=0,
\ee
by means of the change of variables
\be
r= Re^\rv,\ \ \ \phi(r)=e^{-2\rv}\psi(\rv).
\ee
Following \cite{trascfcn}
a solution can be easily written as an expansion for
small $qR$ :
\bea
u(\rv) &=& \sum_{n=0}^{\infty} (qR)^{2n} u_n(\rv), \\
u_0(\rv) &=& \cosh((l+2)\rv), \\
u_n(\rv) &=& -\frac{2}{l+2} \int_0^\rv \cosh(2\rv') \sinh((l+2)(\rv-\rv'))
u_{n-1}(\rv') d\rv'~.
\eea
 A Floquet solution, namely one satisfying
\be
M_\nu(\rv+i\pi) = e^{i\pi\nu} M_\nu(\rv),
\ee
can be obtained from $u(\rv)$ as
\be
M_\nu(\rv) = u(\rv) - e^{i\pi\nu} u(\rv+i\pi),
\ee
where the Floquet exponent $\nu$ turns out to be
\be
\nu = \frac{1}{\pi} \arccos(u(i\pi)).
\ee
This solution can be expanded as
\be \label{mathieusol}
 M_\nu(\rv)= \sum_{n=-\infty}^{\infty} a_{n} e^{(2n+\nu)\rv}~,
\ee
where the coefficients $a_{n}$ can be easily identified from the
expansion
in powers of $qR$ given above. A linearly independent solution to
(\ref{mathieueqn}) is given by
\be 
 M_{-\nu}(\rv)\equiv \sum_{n=-\infty}^{\infty} a_{-n} e^{(2n-\nu)\rv}~,
\ee
where the $a_{n}$ are the \emph{same} coefficients as in
(\ref{mathieusol}).

For ease of comparison, we note that
the Floquet solution $J(\nu,\rv)$ 
given in \cite{gh}
is proportional to $M_\nu$. The former is defined as
\be
J(\nu,\rv)=\sum_{n=-\infty}^{\infty} (-)^{n}\varphi(n+{\textstyle
{1\over 2}}\nu)
  e^{(2n+\nu)\rv}~,
\ee
so the two solutions are related through
\be
 a_{0}J(\nu,\rv) =\varphi(\nu/2)M_\nu(\rv) \quad \Longrightarrow \quad 
(-)^{n}\frac{\varphi(n+\nu/2)}{\varphi(\nu/2)} =
\frac{a_{n}}{a_{0}}~.
\ee
$M_{\nu}$ is similarly related to the  Floquet solution $Me_{\nu}$ 
defined in \cite{mmlz}.

An alternative expansion in terms
of Bessel functions which uses the same coefficients is \cite{gh,mmlz}
\be
 M_\nu(\rv) = \frac{1}{\varphi(\nu/2)}
     \sum_{n=-\infty}^{\infty} (-)^n {a_{n}}
J_n(qRe^{-\rv})J_{n+\nu}(qRe^\rv)~.
\ee
Evaluating this equation at $\rv=0$ and dividing it
by the corresponding equation for $M_{-\nu}(0)=M_\nu(0)$,
the important quantity
$\chi$ follows as
\be
\chi \equiv \frac{\varphi(-\nu/2)}{\varphi(\nu/2)} =
     { { \sum_{n=-\infty}^{\infty} (-)^n a_{-n} J_n(qR)J_{n-\nu}(qR)}
       \over{\sum_{n=-\infty}^{\infty} (-)^n a_{n}
J_n(qR)J_{n+\nu}(qR)}}~.
\ee
The above expression contains only integer powers of $qR$ except for
the factors $(qR)^\nu$ and $(qR)^{-\nu}$
arising from the Bessel functions
$J_{n\pm\nu}(qR)$.
It is possible then to write $\chi$ as
\be
\chi = (qR)^{-2\nu} \tilde{\chi}((qR)^2) = e^{-2\nu\ln(qR)+ i \alpha}
\ee
where $\tilde{\chi}$ (or $\alpha$) can be expressed as a series expansion
in integer powers of $(qR)^2$.
For generic $l>0$ one obtains (following \cite{mmlz})
\bea
\nu &=& l+2 - \frac{(qR)^4}{4(l+1)(l+2)(l+3)} + \cdots \\
\chi &=& (-)^{l+1} \frac{l!(l+1)!}{(l+3)} \frac{2^{2l+2}}{(qR)^{2l}}
 \left[1+\frac{q^4}{4(l+1)(l+2)(l+3)} \ln (qR)^2\right] + \cdots
\eea
which are useful expressions when computing the AdS limit.
In the case of $l=0$ one obtains
\bea
\nu &=& 2+i\mu~, \label{eq:nuAppr} \\
\chi&=&e^{-2i\mu\ln\left[(qR)^2e^\gamma\right]}
\left[-\frac{2}{3}-\frac{1}{3}i\sqrt{5}+(-\frac{49}{1152}
      +\frac{49}{2880}i\sqrt{5})(qR)^4 \right.\nonumber\\
&&\left.+(\frac{18853}{39813120}i\sqrt{5}+\frac{12527}{9953280}
)(qR)^8+\cO((qR)^{12})\right]~, \\
\mu &=& -\frac{\sqrt{5}}{48} (qR)^4+\frac{7}{138240}\sqrt{5}(qR)^8
 +\frac{11851}{637009920}\sqrt{5}(qR)^{12} + \cO((qR)^{16})~, 
\eea
where $\gamma\approx 0.5772$ is Euler's constant.

In the text we also used the solutions \cite{gh} 
\begin{equation}
     H^{(1,2)}(\nu,\rho) = \sum_{n=-\infty}^{\infty} 
     (-)^n {a_{n} \over a_{0}}
     J_{n}(qR e^{-\rho}) H^{(1,2)}_{n+\nu}(qR e^\rho),
     \label{eq:Asymp}
 \end{equation}
where $H^{(i)}_{n+\nu}$ are Hankel functions of order
$n+\nu$. The Mathieu functions $H^{(1,2)}(\nu,\rho)$
are of particular interest
because they behave asymptotically as purely
ingoing or outgoing waves.

\section*{Appendix B}

In this appendix we derive Eq.~(\ref{delta2lexp}). To this end it
is useful
to first summarize some formulas involving the spherical harmonics
on $\bS^5$,  following \cite{lmrs}.
In the following we denote coordinates parallel
and orthogonal to the brane as $x^{\mu}$ and
$\vec{y}$, respectively. Besides we write $\vec{y}=r \hat{y}$, with
$\hat{y}$ a unit vector.
 Orthonormal spherical harmonics on $\bS^5$ are then defined through
\be \label{spharm}
 Y_{l\vm} = \frac{1}{\sqrt{\cNl}} C^{i_1\ldots i_l}_{l\vm}
\hat{y}_{i_1}\ldots \hat{y}_{i_l}~,
\ee
with
\bea
\cNl &=& \frac{\pi^3}{2^{l-1}(l+1)(l+2)}~, \\
\int d^{5}\Omega Y_{l\vm} Y^*_{l'\vm'} &=& \delta_{ll'}\delta_{\vm\vm'}~,\\
\sum_{i_1\ldots i_l}
C_{l\vm}^{i_1\ldots i_l}C_{l\vm'}^{i_1\ldots i_l}&=&\delta_{\vm\vm'}~.
\eea
The coefficients $C^{i_1\ldots i_l}_{l\vm}$ take the symmetric traceless
part
of the product of the unit vectors $\hat{y}$. For a wave incident along
$y_9$ it is important to compute
$\sum_{\vm} C^{9\ldots 9}_{l\vm} C^{9\ldots 9}_{l\vm}$
This calculation was performed in \cite{ktvr} using other coefficients
that can be defined by
\be
C^{i_1\ldots i_l}_{p_1\ldots p_l} = \sum_{\vm}
C_{l\vm}^{i_1\ldots i_l}
C_{l\vm}^{p_1\ldots p_l}~.
\ee
Translating to the notation used here, the result in \cite{ktvr} reads
\bea
\sum_{\vm} C^{9\ldots 9}_{l\vm} C^{9\ldots 9}_{l\vm} &=&
\sum_{i_1\ldots i_l}\sum_{\vm} C^{9\ldots 9}_{l\vm} C_{l\vm}^{i_1\ldots i_l}
\sum_{\vm'} C^{9\ldots 9}_{l\vm'} C_{l\vm'}^{i_1\ldots i_l}
= \sum_{i_1\ldots i_l}
 C^{9\ldots 9}_{i_1\ldots i_l} C^{9\ldots 9}_{i_1\ldots i_l} \nonumber\\
&=&
\frac{(l+2)(l+3)}{3\cdot 2^{l+1}}~.
\eea

Now we proceed to derive Eq.~(\ref{delta2lexp}).
When the interaction is
\be
 S_{\mathrm{int}} = \int d^4x
  \left.{\sum_{{i_{1\dots l}=5}}^9}
  \partial_{i_1\ldots i_l} \phi(x,y) \right|_{y=0}
  C^{i_1\ldots i_l}_{l\vm} \cO^{l\vm}~,
\ee
the analog of (\ref{propexp}) turns out to be
\bea \label{propexpl}
 G(r,r',\hy,\hy') &=& G_0(r,\hy;r',\hy')  \\
&&+ \left.\partial_{i_1\ldots i_l} G_0(r,\hy;\vy_1)\right|_{\vy_1=0}
  \sum_{l\vm} C^{i_1\ldots i_l}_{l\vm} \Delta_2^{(l)}
C^{j_1\ldots j_{l}}_{l\vm} \left.\partial_{j_1\ldots j_l}
G_0(r',\hy';\vy_2)\right|_{\vy_2=0}~. \nonumber
\eea
The flat space propagator $G_0$ can be expanded as
\be
G_0(r,\hy;r',\hy')  = -i \frac{\pi}{2} \frac{1}{(rr')^2} \sum_{l\vm}
J_{l+2}(qr_<) H^{(1)}_{l+2}(qr_>) Y_{l\vm}(\hy) Y_{l\vm}(\hy')
\ee
where $r_{<}$ ($r_{>}$) is the smaller (larger) of $r$ and $r'$.
The Bessel function $J_{l+2}$ ensures that the propagator is well behaved at
$r=0$
and the Hankel function $H^{(1)}_{l+2}$ that the flux at infinity is
outgoing.
{}From the previous equation and (\ref{spharm}) one can compute
\be \label{derG0}
\left.\partial_{i_1\ldots i_l} G_0(r,\hy;\vy)\right|_{\vy=0}
 = -i\frac{\pi}{2} \frac{l!}{(l+2)!} \frac{q^{l+2}}{2^{l+2}}\frac{1}{r^2}
H_{l+2}(qr) \frac{1}{\sqrt{\cNl}}
C^{i_1\ldots i_l}_{l\vm} Y_{l\vm}(\hy) .
\ee
Expanding the full propagator in spherical harmonics,
\be
G(r,\hy;r',\hy')  = \sum_{l\vm} G^{(l)}(r,r') Y_{l\vm}(\hy)
Y_{l\vm}(\hy')~,
\ee
it follows that the coefficients $G^{(l)}(r,r')$
satisfy
\be
\left[ \frac{d^2}{dr^2}+\frac{5}{r}\frac{d}{dr}-\frac{l(l+4)}{r^2} +
q^2 \left(1+\frac{R^4}{r^4}\right)\right]G^{(l)}(r,r')=
\frac{\delta(r-r')}{\pi^3 r^5},
\ee
As in Section \ref{2ptsec}, $G^{(l)}(r,r')$ must reduce to a
purely ingoing
wave at the horizon and a purely outgoing wave at infinity.
Again using the notation of Eq.~(\ref{prop}), the solution is
\be
G^{(l)}(r,r')
       ={1\over w_{32}}\phi_{2}(r_{>})\left[\phi_{3}(r_{<})
       +{B\over A}\phi_{2}(r_{<})\right],
\ee
Expanding for large $r$,$r'$ and using the values of $A$ and $B$ from
(\ref{AB}) the difference between the full and flat space propagators
follows as ($r,r'\rightarrow\infty$):
\be
G^{(l)}(r,r') - G_0^{(l)}(r,r') \simeq \frac{i\pi}{4}\frac{2}{\pi q}
\frac{1}{(rr')^{\frac{5}{2}}}(-)^l e^{iq(r+r')}
\left[(-)^l\frac{\chi_l-\frac{1}{\chi_l}}{\eta_l\chi_l-\frac{1}{\eta_l\chi_l
}}-1
\right]~.
\ee
Finally, inserting the last equation and (\ref{derG0}) into the expansion
(\ref{propexpl}), the propagator $\Delta_2^{(l)}$ is found to be
\be
\Delta_2^{(l)} = 2^{l+5} \pi^2(l+1)(l+2)q^{-2l-4}i
\left[(-)^l\frac{\chi_l-\frac{1}{\chi_l}}{\eta_l\chi_l-\frac{1}{\eta_l\chi_l
}}-1
\right]~.
\ee


\begin{thebibliography}{99}

\bibitem{malda}
J.~Maldacena, ``The Large $N$ Limit of Superconformal Field Theories
and
Supergravity,'' Adv.~Theor. Math.~Phys. {\bf 2} (1998) 231,
{\tt hep-th/9711200}.

\bibitem{gkp}
S.~S. Gubser, I.~R. Klebanov, and A.~M. Polyakov, ``Gauge Theory
Correlators
from Noncritical String Theory,'' Phys.~Lett. {\bf B428} (1998)
105, 
{\tt hep-th/9802109}.

\bibitem{w}
E.~Witten, ``Anti-de Sitter Space and Holography,''
Adv.~Theor. Math.~Phys. {\bf 2} (1998) 253,
{\tt hep-th/9802150}.

\bibitem{magoo}
O.~Aharony, S.~S.~Gubser, J.~Maldacena, H.~Ooguri and Y.~Oz,
``Large N field theories, string theory and gravity,''
{\tt hep-th/9905111}.

\bibitem{polchrr}
J.~Polchinski,
``Dirichlet-Branes and Ramond-Ramond Charges,''
Phys.\ Rev.\ Lett.\  {\bf 75}, 4724 (1995),
{\tt hep-th/9510017}.

\bibitem{3/4}
S.~S.~Gubser, I.~R.~Klebanov and A.~W.~Peet,
``Entropy and Temperature of Black 3-Branes,''
Phys.\ Rev.\  {\bf D54}, 3915 (1996),
{\tt hep-th/9602135}.

\bibitem{klebabs}
I.~R.~Klebanov,
``World-volume approach to absorption by non-dilatonic branes,''
Nucl.\ Phys.\  {\bf B496}, 231 (1997),
{\tt hep-th/9702076}.

\bibitem{gkt}
S.~S.~Gubser, I.~R.~Klebanov and A.~A.~Tseytlin,
``String theory and classical absorption by three-branes,''
Nucl.\ Phys.\  {\bf B499}, 217 (1997),
{\tt hep-th/9703040}.

\bibitem{gk}
S.~S.~Gubser and I.~R.~Klebanov,
``Absorption by branes and Schwinger terms in the world volume theory,''
Phys.\ Lett.\  {\bf B413}, 41 (1997),
{\tt hep-th/9708005}.

\bibitem{ghkk}
S.~S.~Gubser, A.~Hashimoto, I.~R.~Klebanov and M.~Krasnitz,
``Scalar absorption and the breaking of the world volume conformal  
invariance,''
Nucl.\ Phys.\  {\bf B526}, 393 (1998),
{\tt hep-th/9803023}.

\bibitem{gh}
S.~S.~Gubser and A.~Hashimoto,
``Exact absorption probabilities for the D3-brane,''
Commun.\ Math.\ Phys.\  {\bf 203} (1999) 325,
{\tt hep-th/9805140}.

\bibitem{intri}
K.~Intriligator,
``Maximally supersymmetric RG flows and AdS duality,''
{\tt hep-th/9909082}.


\bibitem{taylorvr2}
W.~I.~Taylor and M.~Van Raamsdonk,
``Multiple Dp-branes in weak background fields,''
hep-th/9910052.

\bibitem{myers}
R.~C.~Myers,
``Dielectric-branes,''
JHEP {\bf 9912}, 022 (1999),
{\tt hep-th/9910053}.

\bibitem{dscatt}
A.~Hashimoto and I.~R.~Klebanov,
``Scattering of strings from D-branes,''
Nucl.\ Phys.\ Proc.\ Suppl.\  {\bf 55B}, 118 (1997),
{\tt hep-th/9611214}.

\bibitem{dealwis}
S.~P.~de Alwis,
``Supergravity, the DBI action and black hole physics,''
Phys.\ Lett.\  {\bf B435}, 31 (1998),
{\tt hep-th/9804019}.

\bibitem{susswi}
L.~Susskind and E.~Witten, ``The Holographic Bound in Anti-de Sitter
Space,''
{\tt hep-th/9805114}.

\bibitem{pp}
A.~W.~Peet and J.~Polchinski,
``UV/IR relations in AdS dynamics,''
Phys.\ Rev.\  {\bf D59}, 065011 (1999),
{\tt hep-th/9809022}.

\bibitem{hashimoto}
A.~Hashimoto,
``Holographic description of D3-branes in flat space,''
Phys.\ Rev.\  {\bf D60}, 127902 (1999),
{\tt hep-th/9903227}.

\bibitem{costa}
M.~S.~Costa,
``Absorption by double-centered D3-branes and the 
Coulomb branch of $\cN = 4$  SYM theory,''
{\tt hep-th/9912073}.

\bibitem{costa2}
M.~S.~Costa,
``A Test of the AdS/CFT Duality on the Coulomb Branch,''
{\tt hep-th/0003289}.

\bibitem{tseytlin}
A.~A.~Tseytlin,
``Born-Infeld action, supersymmetry and string theory,''
{\tt hep-th/9908105}.

\bibitem{krvn}
H.~J.~Kim, L.~J.~Romans and P.~van~Nieuwenhuizen, 
``The mass spectrum
of chiral $\cN = 2$ $D = 10$ supergravity on $\bS^5$,''
Phys.\ Rev.\  {\bf D32}, 389  (1985).

\bibitem{ferrara2}
S.~Ferrara, M.~A.~Lled\'o and A.~Zaffaroni,
``Born-Infeld corrections to D3 brane action in AdS$_{5}\times\bS^{5}$
and $\cN = 4$, $d = 4$ primary superfields,''
Phys.\ Rev.\  {\bf D58}, 105029 (1998),
{\tt hep-th/9805082}.

\bibitem{lt}
H.~Liu and A.~A.~Tseytlin,
``Dilaton-fixed scalar correlators and AdS$_{5}\times\bS^{5}$-SYM  
correspondence,''
JHEP {\bf 9910}, 003 (1999),
{\tt hep-th/9906151}.

\bibitem{mmlz}
R.~Manvelian, H.~J.~M\"uller-Kirsten, J.~Q.~Liang and Y.~Zhang,
``Absorption cross-section of scalar field in supergravity background,''
{\tt hep-th/0001179}.

\bibitem{gubser}
S.~S.~Gubser, ``Can the effective string see higher partial waves?,''
Phys.\ Rev.\  {\bf D56}, 4984 (1997),
{\tt hep-th/9704195}.

\bibitem{clvp}
M.~Cveti\v{c}, H.~L\"u and J.~F.~V\a'azquez-Poritz,
``Absorption by extremal D3-branes,''
{\tt hep-th/0002128}.

\bibitem{verlinde}
H.~Verlinde,
``Holography and compactification,''
{\tt hep-th/9906182}.

\bibitem{kv}
J.~Khoury and H.~Verlinde,
``On open/closed string duality,''
{\tt hep-th/0001056}.

\bibitem{pt}
V.~Periwal and {\O}.~Tafjord,
``A finite cutoff on the string world sheet?,''
Phys.\ Rev.\  {\bf D60}, 046004 (1999),
{\tt hep-th/9803195}; \\
{\O}.~Tafjord,
``The Dynamics of D-branes in String Theory,''
  Princeton University Ph.D. Thesis (1999).

\bibitem{park}
I.~Y.~Park,
``Fundamental vs. solitonic description of D3 branes,''
Phys.\ Lett.\  {\bf B468}, 213 (1999),
{\tt hep-th/9907142}.

\bibitem{cm}
N.~R.~Constable and R.~C.~Myers,
``Exotic scalar states in the AdS/CFT correspondence,''
JHEP {\bf 9911}, 020 (1999)
{\tt hep-th/9905081}.

\bibitem{thooft}
G.~'t Hooft,
``Dimensional reduction in quantum gravity,''
{\tt gr-qc/9310026}.

\bibitem{susskind}
L.~Susskind,
``The World as a hologram,''
J.\ Math.\ Phys.\  {\bf 36}, 6377 (1995),
{\tt hep-th/9409089}.

\bibitem{abks}
O.~Aharony, M.~Berkooz, D.~Kutasov and N.~Seiberg,
``Linear dilatons, NS5-branes and holography,''
JHEP {\bf 9810}, 004 (1998).
{\tt hep-th/9808149}.

\bibitem{ms}
S.~Minwalla and N.~Seiberg,
``Comments on the IIA NS5-brane,''
JHEP {\bf 9906}, 007 (1999)
{\tt hep-th/9904142}.

\bibitem{mr}
J.~M.~Maldacena and J.~G.~Russo,
``Large N limit of non-commutative gauge theories,''
JHEP {\bf 9909}, 025 (1999),
{\tt hep-th/9908134}.


\bibitem{bgl}
V.~Balasubramanian, S.~B.~Giddings and A.~Lawrence,
``What do CFTs tell us about anti-de Sitter spacetimes?,''
JHEP {\bf 9903}, 001 (1999),
{\tt hep-th/9902052}.

\bibitem{gidd}
S.~B.~Giddings,
``The boundary S-matrix and the AdS to CFT dictionary,''
Phys.\ Rev.\ Lett.\  {\bf 83}, 2707 (1999),
{\tt hep-th/9903048}.

\bibitem{ktvr}
I.~R.~Klebanov, W.~Taylor IV, and M.~Van~Raamsdonk,
``Absorption of Dilaton Partial Waves by D3-Branes,''
{\tt hep-th/9905174}.


\bibitem{dingle}
R.~B.~Dingle,
\emph{Asymptotic Expansions: Their Derivation and Interpretation},
Academic Press, London (1973).

\bibitem{gr}
I.~S.~Gradshteyn and I.~M.~Ryzhik, 
\emph{Table of Integrals, Series, and Products}, Fifth Edition,
Academic Press, London (1994), p. 973, formula 8.451.5.


\bibitem{bklt}
V.~Balasubramanian, P.~Kraus, A.~Lawrence and S.~P.~Trivedi,
``Holographic probes of anti-de Sitter space-times,''
Phys.\ Rev.\  {\bf D59}, 104021 (1999)
{\tt hep-th/9808017}.

\bibitem{dkk1}
U.~H.~Danielsson, E.~Keski-Vakkuri and M.~Kruczenski,
``Vacua, propagators, and holographic probes in AdS/CFT,''
JHEP {\bf 9901}, 002 (1999),
{\tt hep-th/9812007}.

\bibitem{cg}
C.~G.~Callan and A.~G\"uijosa, 
``Undulating Strings and Gauge Theory Waves,''
Nucl.\ Phys.\  {\bf B565}, 157 (2000),
{\tt hep-th/9906153}.


\bibitem{fmmr}
D.~Z.~Freedman, S.~D.~Mathur, A.~Matusis and L.~Rastelli,
``Correlation functions in the CFT$_d$/AdS$_{d+1}$
correspondence,''
Nucl.\ Phys.\  {\bf B546}, 96 (1999),
{\tt hep-th/9804058}.


\bibitem{reyee}
S.-J.~Rey, J.~Yee,
``Macroscopic Strings as Heavy Quarks of Large $N$ Gauge Theory and
Anti-de
Sitter Supergravity,'' 
{\tt hep-th/9803001}.

\bibitem{juanwilson}
J.~Maldacena, ``Wilson Loops in Large $N$ Field Theories,''
Phys.~Rev.~Lett. {\bf 80} (1998) 4859,
{\tt hep-th/9803002}.

\bibitem{dgo}
N.~Drukker, D.~J.~Gross and H.~Ooguri,
``Wilson loops and minimal surfaces,''
Phys.\ Rev.\  {\bf D60}, 125006 (1999),
{\tt hep-th/9904191}.

\bibitem{sonnenschein}
J.~Sonnenschein,
``What does the string/gauge correspondence teach us about Wilson 
loops?,''
{\tt hep-th/0003032}.

\bibitem{wittenbound}
E.~Witten, ``Bound States of Strings and $p$-Branes,''
\npb{460}{1996}{335}, {\tt hep-th/9510135}.


\bibitem{wbaryon}
E.~Witten, ``Baryons and Branes in Anti de Sitter Space,''
J.~High Energy Phys. {\bf 07} (1998) 006,
{\tt hep-th/9805112}.

\bibitem{groguri}
D.~Gross and H.~Ooguri, ``Aspects of Large N Gauge Theory Dynamics as
seen by String theory,''
Phys.~Rev. {\bf D58} (1998) 106002,
{\tt hep-th/9805129}.

\bibitem{imamura}
Y.~Imamura, ``Supersymmetries and BPS Configurations on Anti-de
Sitter Space,''
Nucl.~Phys. {\bf B537} (1999) 184,
{\tt hep-th/9807179}.

\bibitem{cgs}
C.~G.~Callan, A.~G\"uijosa and K.~G.~Savvidy,
``Baryons and string creation from the fivebrane worldvolume action,''
Nucl.\ Phys.\  {\bf B547}, 127 (1999),
{\tt hep-th/9810092}.

\bibitem{gomis}
B.~Craps, J.~Gomis, D.~Mateos and A.~Van Proeyen,
``BPS solutions of a D5-brane world volume in a D3-brane background 
from  superalgebras,''
JHEP {\bf 9904}, 004 (1999),
{\tt hep-th/9901060}.

\bibitem{gomis2}
J.~Gomis, A.~V.~Ramallo, J.~Simon and P.~K.~Townsend,
``Supersymmetric baryonic branes,''
JHEP {\bf 9911}, 019 (1999)
[hep-th/9907022].

\bibitem{calmal}
C.~Callan and J.~Maldacena, ``Brane Dynamics from the Born-Infeld
Action,''
Nucl.~Phys. {\bf B513} (1998) 198,
{\tt  hep-th/9708147}.

\bibitem{gibb}
G.~Gibbons, ``Born-Infeld Particles and Dirichlet $p$-branes'',
Nucl.~Phys. {\bf B514} (1998) 603,
{\tt  hep-th/9709027}.

\bibitem{camino}
J.~M.~Camino, A.~V.~Ramallo and J.~M.~S\'anchez de Santos,
``Worldvolume dynamics of D-branes in a D-brane background,''
Nucl.\ Phys.\  {\bf B562}, 103 (1999),
{\tt hep-th/9905118}.


\bibitem{hi}
A.~Hashimoto and N.~Itzhaki,
``Non-commutative Yang-Mills and the AdS/CFT correspondence,''
Phys.\ Lett.\  {\bf B465}, 142 (1999),
{\tt hep-th/9907166}.

\bibitem{sw}
N.~Seiberg and E.~Witten,
``String theory and noncommutative geometry,''
JHEP {\bf 9909}, 032 (1999),
{\tt hep-th/9908142}.

\bibitem{mvrs}
S.~Minwalla, M.~Van Raamsdonk and N.~Seiberg,
``Noncommutative perturbative dynamics,''
hep-th/9912072.

\bibitem{vrs}
M.~Van Raamsdonk and N.~Seiberg,
``Comments on noncommutative perturbative dynamics,''
{\tt hep-th/0002186}.

\bibitem{abk}
I.~Y.~Aref'eva, D.~M.~Belov and A.~S.~Koshelev,
``Two-loop diagrams in noncommutative $\phi^{4}_{4}$ theory,''
Phys.\ Lett.\  {\bf B476}, 431 (2000),
{\tt hep-th/9912075}.

\bibitem{abk2}
I.~Y.~Aref'eva, D.~M.~Belov and A.~S.~Koshelev,
``A note on UV/IR for noncommutative complex scalar field,''
hep-th/0001215.

\bibitem{mst}
A.~Matusis, L.~Susskind and N.~Toumbas,
``The IR/UV connection in the non-commutative gauge theories,''
{\tt hep-th/0002075}.

\bibitem{kl}
Y.~Kiem and S.~Lee,
``UV/IR mixing in noncommutative field theory via open string loops,''
{\tt hep-th/0003145}.

\bibitem{abkr}
I.~Y.~Aref'eva, D.~M.~Belov, A.~S.~Koshelev and O.~A.~Rytchkov,
``UV/IR mixing for noncommutative complex scalar field theory.  II: 
(Interaction with gauge fields),''
{\tt hep-th/0003176}.

\bibitem{ad}
O.~Andreev and H.~Dorn,
``Diagrams of noncommutative $\phi^{3}$ theory from string theory,''
{\tt hep-th/0003113}.

\bibitem{bcr}
A.~Bilal, C.~Chu and R.~Russo,
``String theory and noncommutative field theories at one loop,''
{\tt hep-th/0003180}.

\bibitem{gkmrs}
J.~Gomis, M.~Kleban, T.~Mehen, M.~Rangamani and S.~Shenker,
``Noncommutative gauge dynamics from the string worldsheet,''
{\tt hep-th/0003215}.

\bibitem{rr}
A.~Rajaraman and M.~Rozali,
``Noncommutative gauge theory, divergences and closed strings,''
{\tt hep-th/0003227}.

\bibitem{lm}
H.~Liu and J.~Michelson,
``Stretched strings in noncommutative field theory,''
{\tt hep-th/0004013}.

\bibitem{gll}
J.~Gomis, K.~Landsteiner, and E.~L\'opez,
``Non-Relativistic Non-Commutative Field Theory and UV/IR Mixing,''
{\tt hep-th/0004115}.



\bibitem{trascfcn}
A.~Erd\'elyi, ed., 
\emph{Higher Transcendental Functions},
Vol. 3, McGraw-Hill, New York (1955), p.~105.


\bibitem{lmrs}
S.~Lee, S.~Minwalla, M.~Rangamani and N.~Seiberg,
``Three-point functions of chiral operators in D = 4, 
$\cN = 4$ SYM at  large N,''
Adv.\ Theor.\ Math.\ Phys.\  {\bf 2} (1998) 697,
{\tt hep-th/9806074}.


\end{thebibliography}
\end{document}